\documentclass[twocolumn,preprintnumbers,prx]{revtex4}
\usepackage{graphicx}
\usepackage{dcolumn}
\usepackage{bm}
\usepackage{amsmath}
\usepackage{amssymb}
\setcitestyle{super}

\begin{document}

\title{Magnetic excitations in quasi-one dimensional helimagnets:\\Magnon decays and influence of the inter-chain interactions}

\author{Z. Z. Du, H. M. Liu, Y. L. Xie, Q. H. Wang and J. -M. Liu}

\affiliation{Laboratory of Solid State Microstructures and Innovation Center of Advanced Microstructures, Nanjing University, Nanjing 210093, China}

\begin{abstract}
We present a comprehensive study of the magnetic properties of the long-range ordered quasi-one dimensional $J_{1}$-$J_{2}$ systems with a newly developed torque equilibrium spin-wave expansion approach, which can describe the spin Casimir and magnon decay effects in a unified framework. While the framework does not lose the generality, our discussion will be restricted to two representative systems, each of which has only one type of inter-chain coupling ($J_{3}$ or $J_{4}$) and is referred to the $J_{3}$- or $J_{4}$-system respectively. In spite of the long-range spiral order, the dynamical properties of these systems turn out to be highly nontrivial due to the incommensurate noncollinear spin configuration and the strong quantum fluctuation effects enhanced by the frustration and low-dimensionality. Both the systems show prominent spin Casimir effects induced by the vacuum fluctuation of the spin waves and related modification of the ordering vector, Lifshitz point's position and sublattice magnetization. In addition to these static properties, the dynamical behaviors of these systems are also remarkable. Significant and spontaneous magnon decay effects are manifested in the quantum corrections to the excitation spectrum, including the broadening of the spectrum linewidth and downward renormalization of the excitation energy. Furthermore, the excitation spectrum appears to be very sensitive to the types of the inter-chain coupling and manifests three distinct features:
(i) the magnon decay patterns between $J_{3}$- and $J_{4}$-system are very different,
(ii) the renormalized spectrum and the overall decay rate of the $J_{3}$- and $J_{4}$-systems show very different sensitivity to the magnetic anisotropy,
and (iii) there is a nearly flat mode in the renormalized magnon spectrum of the $J_{4}$-system along the X-M direction.
By adjusting the strength of magnetic anisotropy and varying the approximation scheme, it is revealed that these striking distinct features are quite robust and have deep connection with both the spin Casimir and the magnon decay effects.
Thus these special consequences of the inter-chain coupling on the spin wave dynamics may be served as a set of probes for different types of inter-chain couplings in experiments.
At last, to guide experimental measurements such as inelastic neutron scattering in realistic materials and complement our theoretical framework, we develop the analytical theory of the dynamical structure factor within the torque equilibrium formulism and provide the explicit results of the quasi-one dimensional $J_{1}$-$J_{2}$ systems.
\end{abstract}

\maketitle

\section{Introduction}
Frustration and related phenomena in low-dimensional quantum antiferromagnets have been a subject of great interest because of the elusive quest for magnetically disordered phases with highly entangled ground states: quantum spin liquids.~\onlinecite{SL,Wen}
Relentless efforts have been made along the way and a large number of frustrated magnetic systems featuring exotic phases have been discovered and studied.~\onlinecite{SL,OR,QM}
One of the simplest and most investigated frustrated systems is the one-dimensional magnets with a ferromagnetic (FM) nearest-neighbor interaction and an antiferromagnetic (AFM) next-nearest-neighbor interactions, both of which are of intra-chain type.~\onlinecite{QM,SN1,J1,JH1,JH2,JH3,JA1,JA2,Sk1,Sk2}
The majority of theoretical studies on this model are devoted to the ground state phase diagram in the pure one-dimensional case, which includes various exotic quantum phases such as dimer, vector chiral, and spin multi-polar states.~\onlinecite{J1,JH1,JA2}
As a matter of fact, real compounds exhibit besides the significant frustrated intra-chain couplings also relatively weak inter-chain couplings, which can efficiently suppress the quantum fluctuation effect and lead to magnetic long-range order below a Neel temperature $T_{N}$.~\onlinecite{QM}
In this case, with the frustrated intra-chain couplings, the ground state of the system becomes the noncollinear long-range ordered one, which is usually considered to be classical thus has received relatively less attention especially for the excitation spectral properties.
However, there is a significant variety of experimental systems related to this areas of interests,~\onlinecite{Ex1,Ex2,Ex3,Ex4,Ex5,Ex6,MF1,MF2,MF3} in many of which the excitation spectra have been investigated by various experiments including the inelastic neutron and resonant inelastic x-ray scattering.~\onlinecite{Ex1,Ex3,Ex6}

On the other hand, in spite of the magnetic long-range order, the spectral properties of the quasi-one dimensional $J_{1}$-$J_{2}$ systems are actually highly nontrivial due to the strong quantum fluctuation effect enhanced by the frustrated intra-chain couplings and low-dimensionality.
Additionally, there are several important features that make the dynamic properties of these systems significantly rich.
First, the noncollinear nature of the spiral state causes the mixed transverse and longitudinal fluctuations, which can further induce the spontaneous magnon decay effects including the strong renormalization of the spin-wave spectrum and finite magnon lifetime even at zero temperature.~\onlinecite{Oleg,Sasha1,Sasha2,Sasha3}
As a consequence, the magnetic dynamics of the system are qualitatively different from the results obtained within the linear spin-wave theory (LSWT), which is usually used to analyze the experimental results of the excitation spectra.
Second, although the quasi-one dimensionality of these systems indicates an intra-chain coupling dominated magnetic dynamics, the inter-chain coupling can be considerably important as well.~\onlinecite{IC}
However, different from the large intra-chain coupling which is known with reasonable precision, the accurate information about the inter-chain coupling is usually lacking and sometimes which type of the couplings is the leading one remains controversial due to frustration.~\onlinecite{Ex2,AG1,AG2,AG3}
One of the interesting questions is the dependence of the dynamic properties on the inter-chain couplings and whether this dependence is sensitive enough to serve as a probe of different types of inter-chain couplings.
Last but not least, the incommensurate nature of the noncollinear state leads to the spin Casimir effect generated by the zero point fluctuation of the spin wave.
The spin Casimir effect is actually quite general and has different physical meanings in different circumstances, a detailed explanation can be found in our previous work.~\onlinecite{ME}
In general, it represents the macroscopic force or torque that is generated by the vacuum fluctuation of quantum spin systems. In this article, the spin Casimir effect specifically means the quantum fluctuation induced modification of the ordering vector.
In the presence of spin Casimir effect, the conventional spin-wave expansion scheme is plagued with various non-physical singularities and divergences, and the associated spin-wave analysis has to be performed within an alternative formulism.

In this paper we present a comprehensive study of the spin-wave dynamics of the long-range ordered quasi-one dimensional $J_{1}$-$J_{2}$ system with a newly developed spin-wave expansion approach: the torque equilibrium spin wave theory (TESWT).~\onlinecite{ME}
This approach can be considered as an extension of the conventional nonlinear spin-wave theory, in which the spin Casimir effect is treated as a self-consistent manner and the associated expansion results are free from such singularities and divergences.
The details of this expansion formulism have been discussed at length in our previous work,~\onlinecite{ME} which mainly focus on the spin Casimir effect induced singular and divergent problems in the conventional spin-wave expansion scheme and the explicit formulism of the torque equilibrium expansion approach.
In the present work, on the other hand, we extend this expansion formulism to a more realistic multi-parameter case and systematically investigate both the static and dynamic magnetic properties of the systems with the combined spin Casimir and magnon decay effects. It represents a substantial headway towards a comprehensive and advanced theoretical framework of spin wave dynamics in more realistic one-dimensional antiferromagnets with intra-chain and inter-chain interactions.
To investigate the consequence of different inter-chain couplings, we consider two most common types ($J_{3}$ and $J_{4}$) in weakly coupled chain systems (see Fig. 1).
Furthermore, we also study the cases with different strengths of magnetic anisotropy to test the robustness and investigate the physical origin of the distinct features caused by different types of inter-chain couplings.

\begin{figure}
  \includegraphics[width=8.6cm]{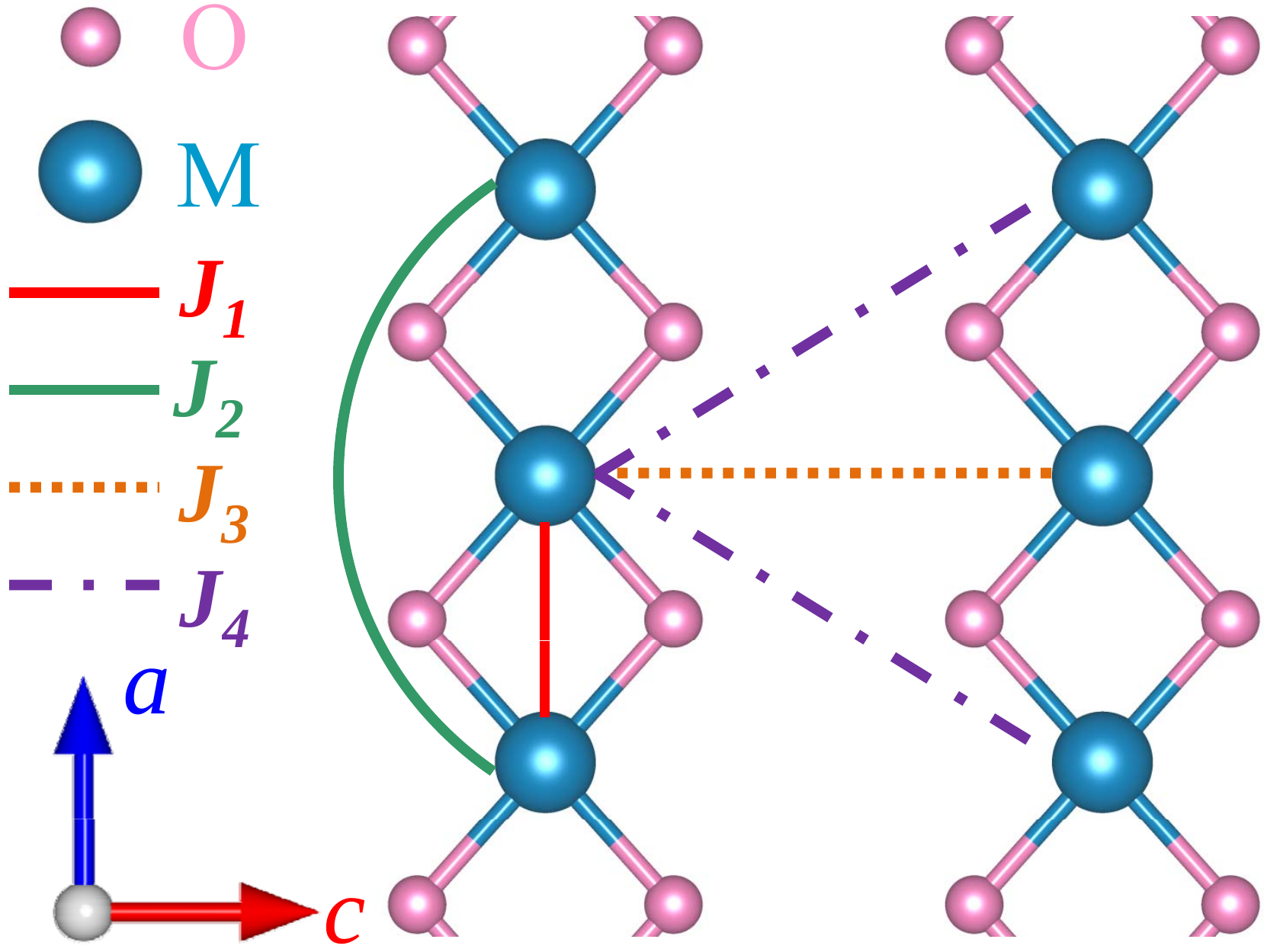}\\
  \caption{Crystallographic structure of coupled edge-shared chain magnetic oxides with main intra- and inter-chain couplings $J_{1}$, $J_{2}$, $J_{3}$ and $J_{4}$ marked by red solid line, green solid arc, orange dotted line, and purple dash dotted line, respectively.}
\end{figure}

In prior to the presentation of our formulation and computations, a brief highlight of the main conclusion is given here. First, the spin Casimir torque induced modification of the ordering vector is obtained by solving the torque equilibrium equation within the one-parameter renormalization approximation. Surprisingly, the magnetic anisotropy induced shift of the FM/spiral quantum phase transition point (Lifshitz point) can be qualitatively manifested in the ordering vector results. As a comparison, standard calculation of this Lifshitz point modification is also performed within both the conventional spin-wave theory (CSWT) and TESWT.
Other than that, the sublattice magnetization in each case is obtained within both spin-wave theories as well.
Besides these static properties, the dynamic magnetic properties are also investigated carefully. The quantum corrections to the excitation spectrum are obtained using both the on-shell and off-shell approximations with the one-loop magnon self-energy and significant magnon decay effects are manifested.
Furthermore, in both the on-shell and off-shell cases the spin-wave spectrum appears to be very sensitive to the types of the inter-chain couplings and the magnetic anisotropy. Interestingly, several remarkable distinct features manifest in the renormalized spectrum of the $J_{3}$- and $J_{4}$-systems, such as the qualitatively different decay patterns, the very dissimilar sensitivities to the magnetic anisotropy and the appearance of a nearly flat mode in the $J_{4}$-systems.
Moreover, these features turn out to be deeply related to both the spin Casimir effect and magnon decay effect, and are expected to be robust. These features may be served as a set of probes for different types of inter-chain couplings.

A surprising but method dependent feature is the "sudden non-decay region" in the off-shell approximated spectrum of the isotropic $J_{4}$-system, which is further analyzed by introducing the poles function and two-magnon density of states.
Our analysis indicates that the appearance of this "sudden non-decay region" is in fact a consequence of the degeneration between the "bonding" and "antibonding" single-magnon states, and thus likely only an artifact of our one-loop approximation.
To verify the influence of the inter-chain couplings on the excitation spectrum and further clarify the methodology associated problems, the spectral function of each system is also obtained, in which the degeneration of single-magnon states is clearly demonstrated.
Furthermore, to guide experimental inelastic neutron scattering measurements and complement our theoretical framework, we develop the analytical theory of the dynamical structure factor $\mathcal{S}(\textbf{k},\varepsilon)$ within the torque equilibrium formulism and provide the explicit results for $\mathcal{S}(\textbf{k},\varepsilon)$ of the quasi-one dimensional $J_{1}$-$J_{2}$ systems.

The rest of this paper is organized as follows.
In Section II we give a short introduction to the quasi-one dimensional $J_{1}$-$J_{2}$ system and the corresponding model Hamiltonian that we shall investigate.
Section III provides a brief review of the nonlinear spin-wave theory and the newly developed torque equilibrium extension.
The spin Casimir torque induced modification of the ordering vector, the magnetic anisotropy induced shift of the quantum Lifshitz point and sublattice magnetization of each system is considered in the Section IV.
Section V is devoted to the calculation of quantum corrections to the spin-wave spectrum within both the on-shell approximation and the off-shell one.
And in Section VI we further investigate the spectral function to verify the excitation spectrum results and clarify the methodology associated problems.
Additionally, we develop the analytical theory of the dynamical structure factor within the torque equilibrium formulism for completeness and present the explicit numerical results in Section VII.
Finally, we draw our discussions and conclusions in Section VIII.

\section{Model Hamiltonian}
As a minimum model, the quasi-one dimensional FM-AFM frustrated $J_{1}$-$J_{2}$ Heisenberg model can be realized in a wide range of materials.~\onlinecite{QM}
One of the best studied family of compounds is the edge-shared chain cuprates, which have attracted much interest recently due to the extremely rich phase diagram with spin multipolar phases observed in high magnetic field.~\onlinecite{J1,JH1,JH2,JH3,Ex3,Ex4,Ex5}
Additionally, this fascinating family can be extended to a more general one, in which Cu is replaced with a general magnetic ion with d orbital and the frustrated $J_{1}$-$J_{2}$ Heisenberg model is realized as follows.
The coupling $J_{1}$ is mediated by the superexchange through the $p$ orbital of the O$^{2-}$ ions and thus strongly depends on the M-O-M bond angle $\theta$.
For $\theta$=$90^{\circ}$, the superexchange process via O$^{2-}$ ions requires the exchange through quasi-orthogonal orbitals on O$^{2-}$, which dictates that the coupling is weakly ferromagnetic.
However, for a structure with $\theta$ distorted away from this high symmetry, the AFM exchange coupling becomes stronger and consequently $J_{1}$ turns from the FM coupling to the AFM coupling at some critical angle $\theta_{c}$.
On the other hand, the coupling $J_{2}$ is mediated by the super-superexchange through M-O-O-M path, and thus is usually AFM with small magnitude that comparable to $J_{1}$.
Usually, this model can also be considered as a spin ladder with frustrated zigzag coupling, as shown in the right panel of Fig. 2.

\begin{figure}
  \includegraphics[width=8cm]{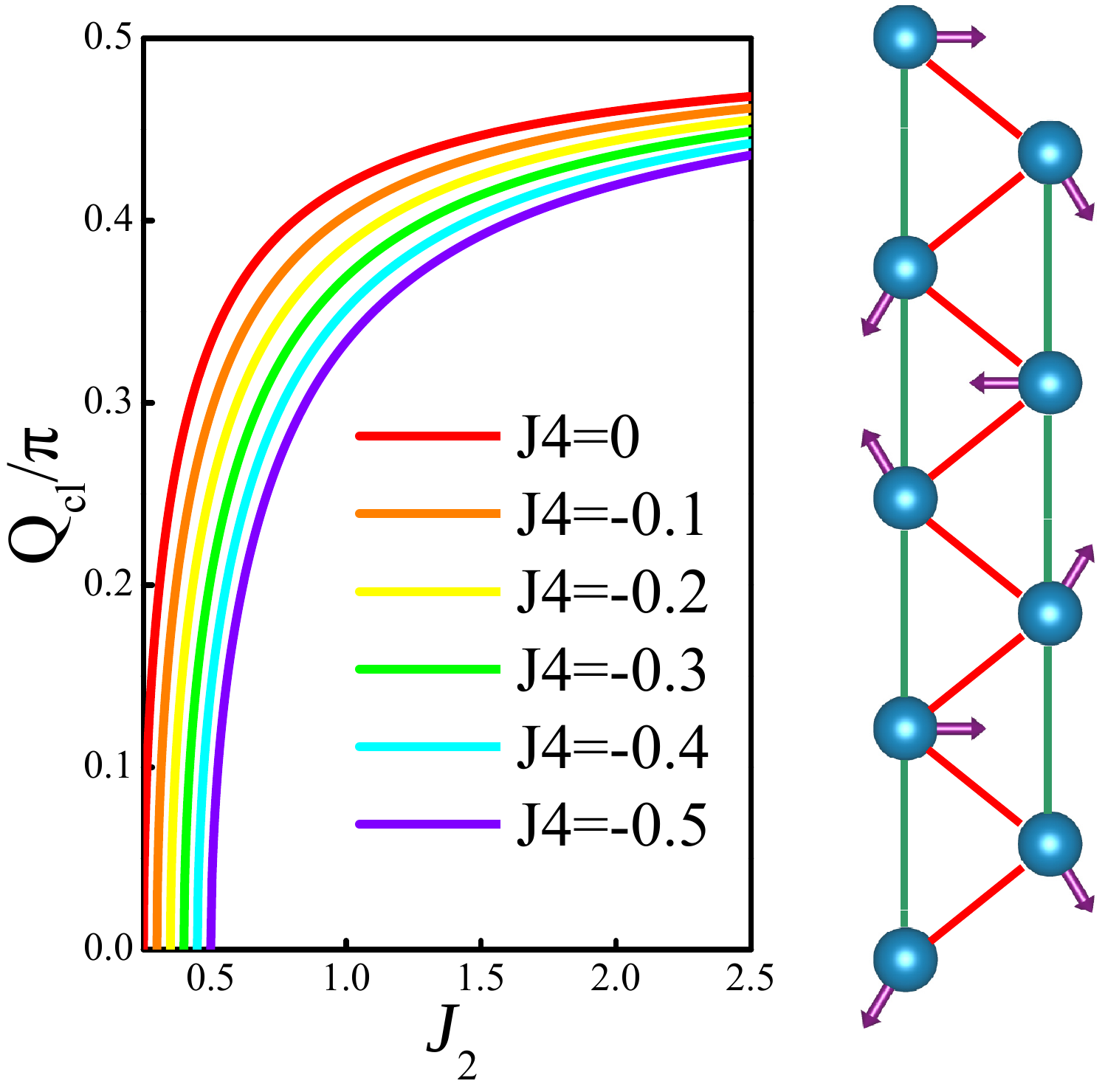}\\
  \caption{Left panel: The classical ordering vector $Q_{cl}$ versus $J_{2}$ with various inter-chain coupling $J_{4}$, where all the parameters are in units of $|J_{1}|$. Right panel: The $J_{1}$-$J_{2}$ chain viewed as a ladder with zigzag coupling and the corresponding spiral spin configuration.}
\end{figure}

In addition to the frustrated intra-chain coupling, an inter-chain coupling is unavoidably present in real materials.
In spite of its weakness in quasi-one dimensional systems, the decisive role of the inter-chain coupling in suppressing the quantum fluctuation is well-known from the Mermin-Wagner-Coleman theorem.~\onlinecite{MWC}
In the absence of frustration, the inter-chain coupling can be determined quite accurately through analyzing T$_{N}$ with Quantum Monte Carlo (QMC) studies.~\onlinecite{QMC}
While in the cases with non-negligible frustration, the QMC is hindered by the so-called sign problem, and a determination of the exchange parameters becomes more difficult.
As a consequence, in a frustrated quasi-one dimensional system, accurate information about the inter-chain couplings is usually lacking.~\onlinecite{Ex2,AG1,AG2,AG3}
Thus, it is instructive to investigate the dependence of dynamic properties on the inter-chain coupling and see whether this dependence is sensitive enough to serve as a probe of different types of inter-chain couplings.

In this paper, we employ a rather simple effective model for the edge-shared chain magnets with the Hamiltonian written as
\begin{equation}
    \hat{\mathcal{H}}=\hat{\mathcal{H}}_{\parallel}+\hat{\mathcal{H}}_{\perp}+\hat{\mathcal{H}}_{\Delta}
\end{equation}
where
\begin{eqnarray}
&&\hat{\mathcal{H}}_{\parallel}=\sum\limits_{i}J_{1}\textbf{S}_{i}\cdot\textbf{S}_{i+a}+J_{2}\textbf{S}_{i}\cdot\textbf{S}_{i+2a}  \nonumber\\
&&\hat{\mathcal{H}}_{\perp}=\sum\limits_{i}J_{3}\textbf{S}_{i}\cdot\textbf{S}_{i+c}+J_{4}(\textbf{S}_{i}\cdot\textbf{S}_{i+a+c}+\textbf{S}_{i}\cdot\textbf{S}_{i+a-c}) \nonumber\\
&&\hat{\mathcal{H}}_{\Delta}=(\Delta-1)\sum\limits_{i}J_{1}S^{b}_{i}S^{b}_{i+a}+J_{2}S^{b}_{i}S^{b}_{i+2a}
\end{eqnarray}
Here $\hat{\mathcal{H}}_{\parallel}$ represents the frustrated $J_{1}$-$J_{2}$ Heisenberg chain with $J_{1}<0$ and $J_{2}>0$.
And $\hat{\mathcal{H}}_{\perp}$ represents the inter-chain coupling, which includes two representative types ($J_{3}$ and $J_{4}$) as demonstrated in Fig. 1.
For the sake of simplicity, both $J_{3}$ and $J_{4}$ are assumed to be ferromagnetic, although an extension to the AFM case is very straightforward.
Other than that, we consider an extension of the frustrated $J_{1}$-$J_{2}$ Heisenberg model to the $XXZ$ model with anisotropy of the easy-plane type $(0\leq\Delta\leq 1)$.
As we shall see, this term can efficiently suppress the magnon decay region despite it gives no contribution to the classical energy and does not affect the cubic magnon vertexes at all.
Under this circumstance, the classical ground state of the system is a spiral state lying in the $a$-$c$ plane and the classical ordering vector $\textbf{Q}_{cl}$=$(Q_{cl},0,0)$ is given by
\begin{equation}
Q_{cl}=\arccos\Big(-\frac{J_{1}+2J_{4}}{4J_{2}}\Big)
\end{equation}
for $|J_{1}+2J_{4}|<4J_{2}$.
For the cases with $|J_{1}+2J_{4}|\geq4J_{2}$, the ground state becomes the FM one.
This result is determined by minimizing the classical ground state energy and plotted in the left panel of Fig. 2, which will be modified once the quantum fluctuation is considered.

For the sake of general interest and highlighting the consequence of different types of inter-chain couplings, we would like to choose the cases with $S$=1/2 as the main focus of this work, which corresponds to the edge-shared chain cuprates. However, we address that our theoretical framework is of generality and can be applied to systems with arbitrary spin length.

\section{Torque Equilibrium Spin Wave Theory}
The spin wave theory or spin-wave expansion approach is based on the assumption that a long range ordered state exists as the ground state and the quantum fluctuation about this classical saddle point is small.~\onlinecite{SWT} Thus, this theory is expected to be less effective for quantum spin systems with low dimensionality or strong frustration, in which quantum fluctuation becomes more important. Surprisingly though, this spin-wave expansion approach is proven to be quite successful in describing the zero-temperature physics of a number of frustrated low-dimensional spin systems.~\onlinecite{QM,Sasha1,Sasha2}
The calculation results show that the anhamonic terms which are usually considered to be weak are actually very important.
Consequently, it is necessary to consider the nonlinear effects of the spin waves in these systems.
However, the conventional spin-wave formulism breaks down generally for those noncollinear ordered systems where the spin Caimir effect causes the shift of the classical saddle point.
To fix this issue, we have developed a modified spin-wave expansion approach named as TESWT in Ref. 31. In this approach, the spin Casimir effect is treated in a self-consistent way, and the spin-wave expansion results are free from singularities and divergences and consistent with previous numerical results.

In the subsequent two subsections we discuss the standard noncollinear spin wave expansion approach and the torque equilibrium formulism. Here we use the basic notations as used in Ref. 31, in which an explicit explanation and derivation of these formulas can be found.

\subsection{Nonlinear spin wave theory}
The standard noncollinear spin wave theory begins by rewriting the spin-$S$ magnetic Hamiltonian for the system from the laboratory frame $(a,b,c)$ to the twisted frame $(x,y,z)$ associated with the classical ground state configuration of the spins as
\begin{eqnarray}
\hat{\mathcal{H}}&=&\sum\limits_{ij}\Big[\Delta J_{ij} S^{y}_{i}S^{y}_{j}+J_{ij}\cos\theta_{ij}(S^{x}_{i}S^{x}_{j}+S^{z}_{i}S^{z}_{j})\nonumber\\
  &&+J_{ij}\sin\theta_{ij}(S^{x}_{i}S^{z}_{j}-S^{z}_{i}S^{x}_{j})\Big]
\end{eqnarray}
with $\theta_{ij}$=$\theta_{j}$-$\theta_{i}$ is the angle between two neighboring spins, which is determined by the ordering vector of the system.
And, without loss of generality, all spins are assumed to lie in the $x$-$z$ plane.
Then, in proceeding with the hermite Holstein-Primakoff
transformation~\onlinecite{HP} of the spin operators into bosons, followed by the Bogolyubov transformation diagonalizing the harmonic part of the bosonic Hamiltonian, we obtained the following effective Hamiltonian:
\begin{eqnarray}
   \hat{\mathcal{H}}_{eff}&=&\sum\limits_{\textbf{k}}\Big[(2S\varepsilon_{\textbf{k}}+\delta\varepsilon_{\textbf{k}})b^{\dagger}_{\textbf{k}}b_{\textbf{k}}-\frac{O_{\textbf{k}}}{2}(b_{\textbf{k}}b_{-\textbf{k}}+b^{\dagger}_{\textbf{k}}b^{\dagger}_{-\textbf{k}})\Big]\nonumber\\
                &&+i\sqrt{2S}\sum\limits_{\textbf{k},\textbf{p}}\Big[\frac{1}{2!}\Gamma_{1}(\textbf{p},\textbf{k}-\textbf{p};\textbf{k})b_{\textbf{k}}b^{\dagger}_{\textbf{k}-\textbf{p}}b^{\dagger}_{\textbf{p}} \nonumber\\
                &&+\frac{1}{3!}\Gamma_{2}(\textbf{p},-\textbf{k}-\textbf{p};\textbf{k})b^{\dagger}_{\textbf{p}}b^{\dagger}_{-\textbf{k}-\textbf{p}}b^{\dagger}_{\textbf{k}}-\textrm{H.c.}\Big]
\end{eqnarray}
Here $\varepsilon_{\textbf{k}}$ represents the harmonic magnon energy spectrum and is given by
\begin{equation}
    \varepsilon_{\textbf{k}}=\sqrt{A^{2}_{\textbf{k}}-B^{2}_{\textbf{k}}}
\end{equation}
with
\begin{eqnarray}
    A_{\textbf{k}}&=&\frac{1}{2}(\Delta J_{\textbf{k}}+\eta_{\textbf{k}}-2J_{\textbf{Q}})\nonumber\\
    B_{\textbf{k}}&=&\frac{1}{2}(\Delta J_{\textbf{k}}-\eta_{\textbf{k}})
\end{eqnarray}
and
\begin{eqnarray}
    J_{\textbf{k}}&=&J_{1}\cos k_{x}+J_{2}\cos2k_{x}+J_{3}\cos k_{y}  \nonumber\\
    &&+2J_{4}\cos k_{x}\cos k_{y} \nonumber\\
    \eta_{\textbf{k}}&=&\frac{1}{2}(J_{\textbf{k}-\textbf{Q}}+J_{\textbf{k}+\textbf{Q}})
\end{eqnarray}

The rest quadratic terms in the effective Hamiltonian come from the Hartree-Fock decoupling of the quartic interaction terms with
\begin{eqnarray}
   \delta\varepsilon_{\textbf{k}}&=&(u^{2}_{\textbf{k}}+v^{2}_{\textbf{k}})\delta A_{\textbf{k}}-2u_{\textbf{k}}v_{\textbf{k}}\delta B_{\textbf{k}} \nonumber\\
    O_{\textbf{k}}&=&(u^{2}_{\textbf{k}}+v^{2}_{\textbf{k}})\delta B_{\textbf{k}}-2u_{\textbf{k}}v_{\textbf{k}}\delta A_{\textbf{k}}
\end{eqnarray}
where $u_{\textbf{k}}$ and $v_{\textbf{k}}$ are the Bogolyubov transformation coefficients, which are under conditions $u^{2}_{\textbf{k}}-v^{2}_{\textbf{k}}=1$,
\begin{equation}
    u^{2}_{\textbf{k}}+v^{2}_{\textbf{k}}=\frac{A_{\textbf{k}}}{\varepsilon_{\textbf{k}}},~~~~2u_{\textbf{k}}v_{\textbf{k}}=\frac{B_{\textbf{k}}}{\varepsilon_{\textbf{k}}}
\end{equation}
and
\begin{eqnarray}
    \delta A_{\textbf{k}}&=&A_{\textbf{k}}+\sum\limits_{\textbf{p}}\frac{1}{\varepsilon_{\textbf{p}}}\Big[A_{\textbf{p}}(A_{\textbf{k}-\textbf{p}}-A_{\textbf{k}}-A_{\textbf{p}}-B_{\textbf{k}-\textbf{p}}) \nonumber\\
    &&+B_{\textbf{p}}(\frac{B_{\textbf{k}}}{2}+B_{\textbf{p}})\Big]\nonumber\\
    \delta B_{\textbf{k}}&=&B_{\textbf{k}}-\sum\limits_{\textbf{p}}\frac{1}{\varepsilon_{\textbf{p}}}\Big[B_{\textbf{p}}(A_{\textbf{k}-\textbf{p}}-\frac{A_{\textbf{k}}}{2}-A_{\textbf{p}}-B_{\textbf{k}-\textbf{p}}) \nonumber\\
    &&+A_{\textbf{p}}(B_{\textbf{k}}+B_{\textbf{p}})\Big]
\end{eqnarray}
The cubic interaction terms which vanish in collinear magnetic systems are given by
\begin{eqnarray}
   \Gamma_{1}(\textbf{1},\textbf{2};\textbf{3})&=&\frac{-1}{2\xi}\Big[\zeta_{\textbf{1}}\kappa_{\textbf{1}}(\gamma_{\textbf{2}}\gamma_{\textbf{3}}+\kappa_{\textbf{2}}\kappa_{\textbf{3}})
                      +\zeta_{\textbf{2}}\kappa_{\textbf{2}}(\gamma_{\textbf{1}}\gamma_{\textbf{3}} \nonumber\\
                      &&+\kappa_{\textbf{1}}\kappa_{\textbf{3}})+\zeta_{\textbf{3}}\kappa_{\textbf{3}}(\gamma_{\textbf{1}}\gamma_{\textbf{2}}-\kappa_{\textbf{1}}\kappa_{\textbf{2}})\Big]
                      \nonumber\\
                       \nonumber\\
   \Gamma_{2}(\textbf{1},\textbf{2};\textbf{3})&=&\frac{1}{2\xi}\Big[\zeta_{\textbf{1}}\kappa_{\textbf{1}}(\gamma_{\textbf{2}}\gamma_{\textbf{3}}-\kappa_{\textbf{2}}\kappa_{\textbf{3}})
                     +\zeta_{\textbf{2}}\kappa_{\textbf{2}}(\gamma_{\textbf{1}}\gamma_{\textbf{3}}
                     \nonumber\\
                     &&-\kappa_{\textbf{1}}\kappa_{\textbf{3}})+\zeta_{\textbf{3}}\kappa_{\textbf{3}}(\gamma_{\textbf{1}}\gamma_{\textbf{2}}-\kappa_{\textbf{1}}\kappa_{\textbf{2}})\Big]
\end{eqnarray}
with
\begin{equation}
\zeta_{\textbf{k}}=\frac{1}{2}(J_{\textbf{k}-\textbf{Q}}-J_{\textbf{k}+\textbf{Q}})
\end{equation}
and
\begin{equation}
   \xi=\sqrt{\varepsilon_{\textbf{1}}\varepsilon_{\textbf{2}}\varepsilon_{\textbf{3}}},~~~
   \kappa_{\textbf{i}}=\sqrt{A_{\textbf{i}}+B_{\textbf{i}}},~~~
   \gamma_{\textbf{i}}=\sqrt{A_{\textbf{i}}-B_{\textbf{i}}}
\end{equation}
where $\textbf{i}\in(\textbf{1},\textbf{2},\textbf{3})$ and $\textbf{1}, \textbf{2}...$ denote $\textbf{k}_{1}, \textbf{k}_{2}...$.

It is obvious that the spin-wave expansion contributes to the corrections of the ground state energy as well. With the vacuum energy modified from $E_{cl}$ to $E_{vac}$, the ordering vector of the system should be determined by
minimizing $E_{vac}$ via $\delta E_{vac}/\delta \textbf{Q}=0$.
The modification of the classical ordering vector is actually a shift of the classical saddle point due to the zero-point fluctuation.
However, the harmonic spin-wave spectrum function $\varepsilon_{k}$ is only well-defined at $\textbf{Q}=\textbf{Q}_{cl}$, and thus the variation is normally treated approximately as an expansion around $\textbf{Q}_{cl}$.
The conventional $1/S$ order expansion result is
\begin{equation}
   \textbf{Q}=\textbf{Q}_{cl}+\textbf{Q}_{1}
\end{equation}
with
\begin{equation}
   \textbf{Q}_{1}=-\frac{1}{2S}\Bigg[\frac{\partial^{2}J_{\textbf{Q}}}{\partial \textbf{Q}^{2}}\Bigg]^{-1}
                \sum\limits_{\textbf{k}}\frac{A_{\textbf{k}}+B_{\textbf{k}}}{\varepsilon_{\textbf{k}}}\cdot\frac{\partial J_{\textbf{k}+\textbf{Q}}}{\partial \textbf{Q}}\Bigg|_{\textbf{Q}_{cl}}
\end{equation}
This result seems reasonable and is usually treated as the new ordering vector of the system. However, this direct expansion is actually divergent at the spiral/Neel Lifshitz point, thus can not be considered as a correction to the classical ordering vector.~\onlinecite{ME}
More than that, this direct expansion procedure further leads to various divergent results and thus invalidates the CSWT.
As a consequence, some modifications have to be made to obtain a consistent spin-wave description of the system.
\subsection{The torque equilibrium formulism}
The torque equilibrium formulism is based on the spin Casimir interpretation of the saddle point shifting problem.
The spin Casimir torque that accounts for the shift of the saddle point is defined as~\onlinecite{ME}
\begin{equation}
   \textbf{T}_{sc}(\textbf{Q})=\sum\limits_{\textbf{k}}\Bigg\langle\Psi_{vac}\Bigg|\frac{\partial\hat{\mathcal{H}}_{sw}}{\partial \textbf{Q}}\Bigg|\Psi_{vac}\Bigg\rangle
\end{equation}
where $|\Psi_{vac}\rangle$ represents the quasi-particle vacuum state and $\hat{\mathcal{H}}_{sw}$ denotes the spin wave Hamiltonian before the Bogoliubov transformation.
Notice that $\textbf{T}_{sc}$ is a function of $\textbf{Q}$ defined on bonds and represents the tendency of modification to the relative orientation of each spin.
The saddle point condition is given by the torque equilibrium condition
\begin{equation}
\textbf{T}_{cl}(\textbf{Q})+\textbf{T}_{sc}(\textbf{Q})=0
\end{equation}
where $\textbf{T}_{cl}(\textbf{Q})$=$\partial E_{cl}/\partial\textbf{Q}$ represents the classical spin torque.

The core ingredient of the torque equilibrium formulism is to map the original spin system to a new spin system that has the same symmetry and set of exchange integrals as the original one. Consequently, this new spin system is nothing but the original one with a different parameter $J_{i}$ denoted as $\widetilde{J_{i}}$.
The new system with $\widetilde{J_{i}}$ has classical ordering vector identical with the modified one in the old system $\widetilde{\textbf{Q}}_{cl}=\textbf{Q}$.
Thus, the old system with shifted saddle point can be described by
\begin{equation}
\hat{\mathcal{H}}_{sw}(J_{i},\textbf{Q})=\widetilde{\mathcal{H}}_{sw}(\widetilde{{J_{i}}},\textbf{Q})+\mathcal{H}_{sw}^{c}
\end{equation}
with
\begin{equation}
\mathcal{H}_{sw}^{c}=\hat{\mathcal{H}}_{sw}(J_{i},\textbf{Q})-\widetilde{\mathcal{H}}_{sw}(\widetilde{{J_{i}}},\textbf{Q})
\end{equation}
Note that $\mathcal{H}_{sw}$ can be written as series of terms with different orders such as
$\hat{\mathcal{H}}_{2}$,$\hat{\mathcal{H}}_{3}$,$\hat{\mathcal{H}}_{4}$ and so on, it is obvious that $\mathcal{H}_{sw}^{c}$ can be recast in the same form as well and its exact expression is fixed with physical renormalization conditions, which regularizes all the divergences as the counter-terms introduced in quantum field theory.
The explicit proof of the divergent cancelation can be found in Ref. 31, and here we only list the final results for latter convenience.
Given that we are only interested in the results at $1/S$ order, the higher order terms such as $\mathcal{H}^{c}_{3}$ and $\mathcal{H}^{c}_{4}$ can be neglected and we obtain
\begin{equation}
    \widetilde{\mathcal{H}}_{sw}=\widetilde{\mathcal{H}}_{2}+\mathcal{H}^{c}_{2}+\widetilde{\mathcal{H}}_{3}+\widetilde{\mathcal{H}}_{4}
\end{equation}
where $\widetilde{R}$ represents $R(\widetilde{J_{i}},\textbf{Q})$ and $R^{c}$=$R-\widetilde{R}$.
It is obvious that $\mathcal{H}^{c}_{2}$ is of the same order as $\widetilde{\mathcal{H}}_{4}$ and thus treated as perturbation.
As a consequence, the spin Casimir torque in the torque equilibrium condition becomes $\widetilde{\textbf{T}}_{sc}(\textbf{Q})$
and the torque equilibrium equation turns into
\begin{equation}
\textbf{T}_{cl}(\textbf{Q})+\widetilde{\textbf{T}}_{sc}(\textbf{Q})=0
\end{equation}

Based on this renormalization condition and treating $\mathcal{H}_{2}^{c}$ as perturbation,
the one-loop torque equilibrium effective Hamiltonian reads
\begin{eqnarray}
   \widetilde{\mathcal{H}}_{eff}&=&\sum\limits_{\textbf{k}}\Bigg\{(2S\widetilde{\varepsilon}_{\textbf{k}}+\delta\widetilde{\varepsilon}_{\textbf{k}})b^{\dagger}_{\textbf{k}}b_{\textbf{k}}
                   -\frac{\widetilde{O}_{\textbf{k}}}{2}(b_{\textbf{k}}b_{-\textbf{k}}+b^{\dagger}_{\textbf{k}}b^{\dagger}_{-\textbf{k}})\nonumber\\
                &&+2S\Big[\varepsilon^{c}_{\textbf{k}}b^{\dagger}_{\textbf{k}}b_{\textbf{k}}
                   -\frac{O^{c}_{\textbf{k}}}{2}(b_{\textbf{k}}b_{-\textbf{k}}+b^{\dagger}_{\textbf{k}}b^{\dagger}_{-\textbf{k}})\Big]\Bigg\}\nonumber\\
                   \nonumber\\
                &&+i\sqrt{2S}\sum\limits_{\textbf{k},\textbf{p}}\Big[\frac{1}{2!}\widetilde{\Gamma}_{1}(\textbf{p},\textbf{k}-\textbf{p};\textbf{k})b_{\textbf{k}}b^{\dagger}_{\textbf{k}-\textbf{p}}b^{\dagger}_{\textbf{p}} \nonumber\\
                &&+\frac{1}{3!}\widetilde{\Gamma}_{2}(\textbf{p},-\textbf{k}-\textbf{p};\textbf{k})b^{\dagger}_{\textbf{p}}b^{\dagger}_{-\textbf{k}-\textbf{p}}b^{\dagger}_{\textbf{k}}-\textrm{H.c.}\Big]
\end{eqnarray}
with
\begin{eqnarray}
    \varepsilon^{c}_{\textbf{k}}&=&(\widetilde{u}^{2}_{\textbf{k}}+\widetilde{v}^{2}_{\textbf{k}})A^{c}_{\textbf{k}}-2\widetilde{u}_{\textbf{k}}\widetilde{v}_{\textbf{k}}B^{c}_{\textbf{k}} \nonumber\\
    O^{c}_{\textbf{k}}&=&(\widetilde{u}^{2}_{\textbf{k}}+\widetilde{v}^{2}_{\textbf{k}})B^{c}_{\textbf{k}}-2\widetilde{u}_{\textbf{k}}\widetilde{v}_{\textbf{k}}A^{c}_{\textbf{k}}
\end{eqnarray}

According to this torque equilibrium effective Hamiltonian and the standard diagrammatic technique for bosons at zero temperature, the bare magnon propagator can be defined as
\begin{equation}
G^{-1}_{0}(\textbf{k},\varepsilon)=\varepsilon-2S\widetilde{\varepsilon}_{\textbf{k}}+i0^{+}
\end{equation}
Different from the standard nonlinear spin-wave expansion results, we obtain besides the two frequency independent Hartree-Fock contributions to the normal and anomalous self-energies
\begin{equation}
\Sigma^{a}_{hf}(\textbf{k})=\delta\widetilde{\varepsilon}_{\textbf{k}},~~~
\Sigma^{b}_{hf}(\textbf{k})=-\widetilde{O}_{\textbf{k}}
\end{equation}
another two frequency independent contributions
\begin{equation}
\Sigma^{a}_{c}(\textbf{k})=2S\varepsilon^{c}_{\textbf{k}},~~~
\Sigma^{b}_{c}(\textbf{k})=-2SO^{c}_{\textbf{k}}
\end{equation}
These two terms are actually of order $O(S^{0})$ in spite of the $2S$ coefficient.
The last but not the least, the most important normal self-energies contributed from the cubic vertexes are
\begin{eqnarray}
\Sigma^{a}_{3}(\textbf{k},\varepsilon)&=&\frac{1}{2}\sum_{\textbf{p}}\frac{|\widetilde{\Gamma}_{1}(\textbf{p};\textbf{k})|^{2}}
{\varepsilon-\widetilde{\varepsilon}_{\textbf{p}}-\widetilde{\varepsilon}_{\textbf{k}-\textbf{p}}+i0^{+}}  \nonumber\\
\Sigma^{b}_{3}(\textbf{k},\varepsilon)&=&-\frac{1}{2}\sum_{\textbf{p}}\frac{|\widetilde{\Gamma}_{2}(\textbf{p};\textbf{k})|^{2}}
{\varepsilon+\widetilde{\varepsilon}_{\textbf{p}}+\widetilde{\varepsilon}_{\textbf{k}+\textbf{p}}-i0^{+}}
\end{eqnarray}
and the anomalous self-energies contributed from the cubic vertexes are
\begin{eqnarray}
\Sigma^{c}_{3}(\textbf{k},\varepsilon)&=&-\frac{1}{2}\sum_{\textbf{p}}
                      \frac{\widetilde{\Gamma}_{1}(-\textbf{k},\textbf{p})\widetilde{\Gamma}_{2}(\textbf{k},\textbf{p})}{\varepsilon+\widetilde{\varepsilon}_{\textbf{p}}+\widetilde{\varepsilon}_{\textbf{k}+\textbf{p}}-i0^{+}} \nonumber\\
\Sigma^{d}_{3}(\textbf{k},\varepsilon)&=&\frac{1}{2}\sum_{\textbf{p}}
                      \frac{\widetilde{\Gamma}_{1}(\textbf{k},\textbf{p})\widetilde{\Gamma}_{2}(-\textbf{k},\textbf{p})}{\varepsilon-\widetilde{\varepsilon}_{\textbf{p}}-\widetilde{\varepsilon}_{\textbf{k}-\textbf{p}}+i0^{+}}
\end{eqnarray}
The diagrammatic representations of the these self-energies can be found in Fig.3 and Fig.5 of Ref. 31.
All these lowest order results provide a basis for the systematic pertuabative calculations of various static and dynamic magnetic properties of the system.

\section{Static properties}
The pure one dimensional $J_{1}$-$J_{2}$ model has been studied theoretically with much success over the last two decades relying on the availability of many exact results and the absence of the size constrain of the density matrix renormalization group (DMRG) calculation.~\onlinecite{QM,SN1,J1,JH1,JH2,JH3,JA1,JA2,Sk1,Sk2}
As a consequence, most of the studies on the quasi-one dimensional models are carried out by perturbating the one dimensional results with weak inter-chain coupling using the bosonization method.~\onlinecite{J1,JH2}
On the contrary, the experimental results of realistic materials indicate that most of the quasi-one dimensional systems are in fact long range ordered at sufficiently low temperature.~\onlinecite{QM,Ex1,Ex6}
Thus, it seems that the spin wave description of the weakly coupled frustrated chain systems is legitimate at least at zero temperature with sufficiently strong inter-chain coupling.
However, to the best of our knowledge, a reliable spin wave analysis of these systems is still lacking because the LSWT results are unreliable for neglecting the strong quantum fluctuation effect and the conventional $1/S$ expansion scheme is plagued with divergent problems due to the incommensurate noncollinear spin configuration.~\onlinecite{ME}
Under this circumstance, the fitting results of the experimental measurements obtained based on the conventional spin wave analysis turn out to be very inaccurate and can lead to controversial conclusions about the corresponding magnetic models.~\onlinecite{Ex1,Ex2,AG1,AG2,AG3}

In the subsequent three subsections, we shall investigate the static properties of the quasi-one dimensional $J_{1}$-$J_{2}$ systems within both the linear approximated CSWT and the TESWT. Our results indicate that the linear approximated TESWT provides a relatively more accurate description and thus may be served as an efficient parameter fitting tool for the incommensurate noncollinear ordered magnetic systems.
The emergence of the spin Casimir torque and corresponding modification of the ordering vector are investigated in the first subsection.
The standard calculation of the magnetic anisotropy induced shift of the FM/spiral Lifshitz point and the sublattice magnetization is performed in the second and last subsections respectively.

\begin{figure*}
  \includegraphics[width=16cm]{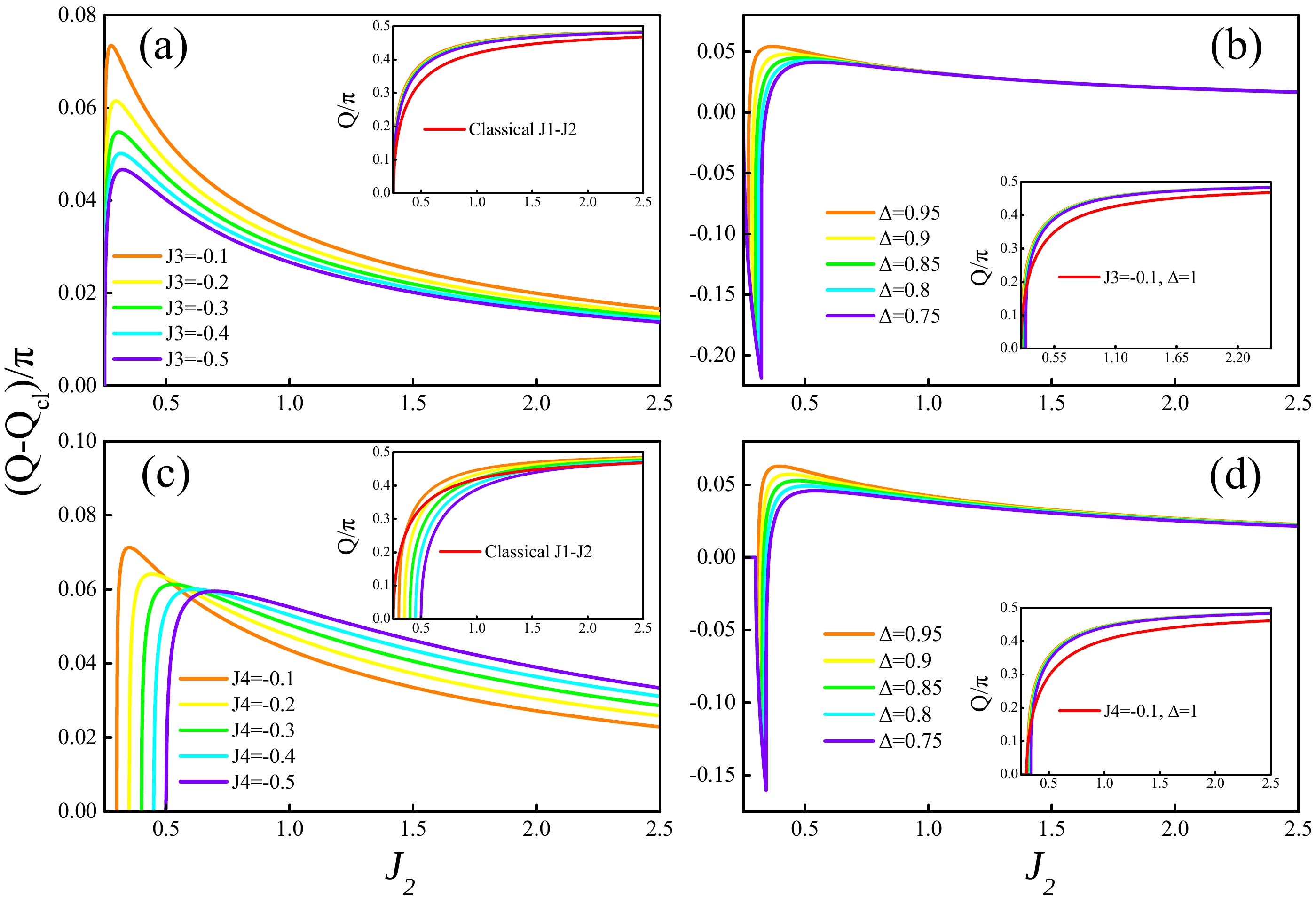}\\
  \caption{The quantum ordering vector $Q$ versus $J_{2}$ in $J_{3}$- and $J_{4}$-systems: (a) isotropic systems with various inter-chain coupling $J_{3}$; (b) anisotropic systems with various anisotropic parameter $\Delta$ and $J_{3}$ fixed to -0.1; (c) isotropic systems with various inter-chain coupling $J_{4}$; (d) anisotropic systems with various anisotropic parameter $\Delta$ and $J_{4}$ fixed to -0.1. Here all the exchange interactions are in units of $|J_{1}|$. }
\end{figure*}

\subsection{Ordering vector}
The quantum fluctuation induced renormalization of the classical ordering vector is a widespread phenomenon in noncollinear ordered antiferromagnets, especially for the systems with incommensurate spin correlation.~\onlinecite{ME,Sk1,CCM}
This modification indicates that the quantum fluctuation can induce a shift of the saddle point due to the spin Casimir effect. To incorporate this effect within the spin wave theory, the spin torque equilibrium condition has to be fulfilled. From our previous definition, the spin Casimir torque in our system can be easily obtained as
\begin{equation}
   \textbf{T}_{sc}(\textbf{Q})
            =\frac{S}{2}\sum\limits_{\textbf{k}}\frac{A_{\textbf{k}}+B_{\textbf{k}}}{\varepsilon_{\textbf{k}}}\cdot\frac{\partial J_{\textbf{k}+\textbf{Q}}}{\partial \textbf{Q}}
\end{equation}
Note that $\textbf{T}_{sc}(\textbf{Q})$ represents the quantum fluctuation induced modification of the classical ordering vector and depends on all the parameters in the spin-wave Hamiltonian.
As a consequence, the quantum fluctuation modified ordering vector $\textbf{Q}$ that is obtained by solving the torque equilibrium equation
\begin{equation}
\frac{\partial J_{\textbf{Q}}}{\partial \textbf{Q}}=-\frac{1}{2S}\sum\limits_{\textbf{k}}\frac{\widetilde{A}_{\textbf{k}}+\widetilde{B}_{\textbf{k}}}{\widetilde{\varepsilon}_{\textbf{k}}}\cdot\frac{\partial \widetilde{J}_{\textbf{k}+\textbf{Q}}}{\partial \textbf{Q}}
\end{equation}
also depends on all the parameters. Based on this observation, in principle, we should consider the renormalization of all these parameters when the torque equilibrium equation is solved. However, as argued in Ref. 31, the renormalization of more than one parameter is tedious and usually unnecessary. Thus, here we only consider the renormalization of the dominant intra-chain parameter $\alpha$=$J_{1}/J_{2}$ and neglect the renormalization of the other small parameters. This is expected to be a good approximation for the quasi-one dimensional cases where the intra-chain coupling is much greater than the inter-chain coupling. Additionally, the validity of this approximation can be easily verified by testing the sensitivity of the solution of the equation ($\widetilde{\alpha}$ and $\textbf{Q}$) to the other parameters.

Other than the verification of our approximation scheme, the dependence of the quantum ordering vector on various parameters is of interest on its own right. To explicitly show the role of each parameter, we consider two representative systems, each of which has only one type of inter-chain coupling. First, we consider the isotropic system with the direct FM inter-chain coupling $J_{3}$. The ordering vector in dependence of $J_{2}$ with different $J_{3}$ is demonstrated in the insert of Fig. 3(a).
It is clearly shown that the ordering vector is drastically modified from it's classical value by the quantum fluctuation effect, which indicates the necessity of including the spin Casimir contribution.
Additionally, the quantum ordering vector $\textbf{Q}$ is mainly dependent of $J_{2}$ and does not show strong sensitivity to the inter-chain coupling and the results with different $J_{3}$ are nearly identical.
As the magnetic anisotropy is introduced, the sensitivity of the quantum ordering vector to $\Delta$ is slightly stronger compared to $J_{3}$ especially near the FM/spiral Lifshitz point as shown in Fig. 3(b). Interestingly, this sensitivity is actually a consequence of the shift of the magnetic anisotropy induced modification of the Lifshitz point.~\onlinecite{JA1,JA2}
This is different from the classical case, where the magnetic anisotropy does not affect the Lifshitz point at all. Furthermore, this is also different from the CSWT description, in which the magnetic anisotropy induced shift of the Lifshitz point can only be obtained by comparing the ground state energy as we shall discuss in the next subsection.
Actually, it is a special property of the TESWT that the ordering vector results can indicate the shift of the phase boundary, such as the case in Ref. 31 the ordering vector can manifest a possible quantum order by disorder (QObD) effect.~\onlinecite{OBD1,OBD2}
However, once $J_{2}$ is away from the Lifshitz point, this sensitivity is not obvious.
Thus, our one parameter renormalization scheme turns out to be a good approximation for the $J_{3}$-systems if not too close to the Lifshitz point.

Next we discuss the systems with the crossed FM inter-chain coupling $J_{4}$.
The situation here is more complicated than that in the $J_{3}$-systems because $J_{4}$ can induce modification of the ordering vector already at the classical level, as shown in Fig. 2.
As a consequence, all the results we obtain on this system have mixed classical and quantum contributions. To eliminate the classical effect and show the quantum part more clearly, we also plot the difference between the quantum ordering vector $\textbf{Q}$ and its classical counterpart $\textbf{Q}_{cl}$.
As shown in Fig. 3(c), the isotropic $J_{4}$-systems show relatively strong sensitivity to the inter-chain coupling than the $J_{3}$-systems.
Nevertheless, the quantum corrections to the classical ordering vector can still be considered to be insensitive to the inter-chain coupling for cases of $J_{2}$ away from the Lifshitz point. On the other hand, the effect of the magnetic anisotropy on the $J_{4}$-systems shows very similar behavior to that on the $J_{3}$-systems away from the Lifshitz point.
However, the behavior around the Lifshitz point is very different due to the mixed classical and quantum contribution to the modification of the Lifshitz point.
To the convenience of comparison, we plot in Fig.4 the Lifshitz points' position of each anisotropic $J_{3}$- and $J_{4}$-system that read from Fig. 3(b) and (d) to demonstrate the effect of the magnetic anisotropy on the modification of the Lifshitz point.

Combined with the results of these two representative systems, it is obvious that our one-parameter renormalization scheme can be considered as a good approximation for both situations, at least in regions away from the Lifshitz point.
Moreover, our TESWT results show good consistency with previous numerical results obtained with the coupled cluster method,~\onlinecite{CCM} different from the classical prediction but less developed than one would expect from the pure one-dimensional results.~\onlinecite{Sk1,Sk2}
This is exactly the situation that one usually encounters in the experimental parameter fitting processes for quasi-one dimensional incommensurate noncollinear ordered magnets: the LSWT results are too classical while the pure one dimensional DMRG results seem to be too quantum.~\onlinecite{Ex1,AG3}
Other than that, our TESWT predictions become identical with the CSWT results once the spin Casimir effect is absent.
Thus it seems that our TESWT can provide a rather accurate prediction of the ordering vector in a general sense.

\begin{figure}
  \includegraphics[width=8.6cm]{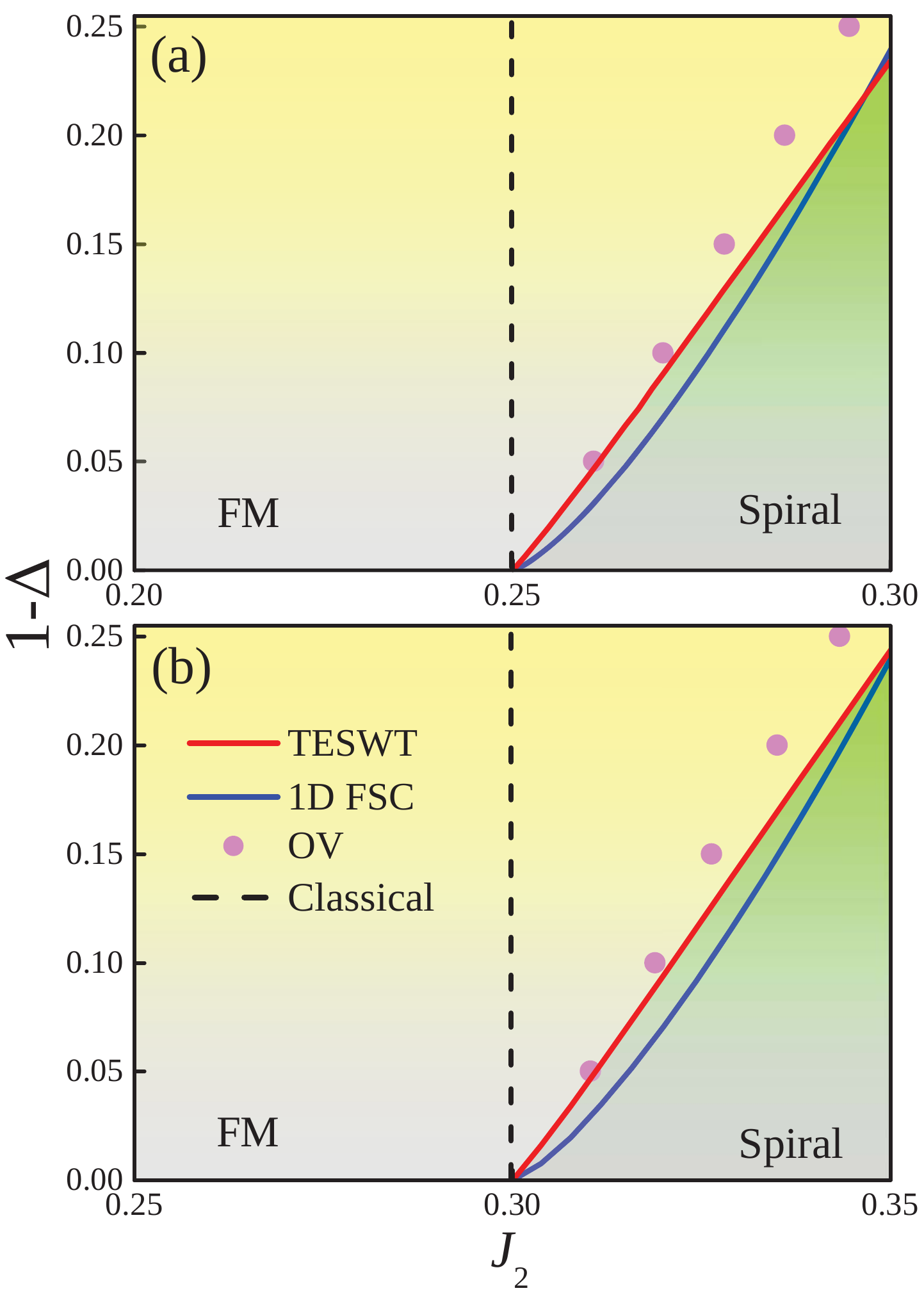}\\
  \caption{The phase diagram with anisotropic parameter $\Delta$ in (a) $J_{3}$-systems with $J_{3}$=-0.1 and (b) $J_{4}$-systems with $J_{4}$=-0.1. The phase boundary obtained through classical and TESWT are marked by black dash line and red solid line respectively. The blue solid line is the fitting line to the finite size calculation (FSC) results of the one dimensional frustrated zigzag $XXZ$ model. And the pink dots are transition points obtained from the ordering vector (OV) data. Here all the exchange interactions are in units of $|J_{1}|$.}
\end{figure}

\subsection{FM/spiral Lifshitz point}
In both the quantum and classical models of the FM-AFM frustrated $J_{1}$-$J_{2}$ chain systems, the presence of a Lifshitz point at $J_{2}$=$|J_{1}|/4$ is well known.~\onlinecite{QM}
In the classical case, this Lifshitz point describes a zero temperature transition from the ferromagnetic state to spiral state, whose position is purely determined by the exchange couplings thus independent on the magnetic anisotropy.
As the quantum fluctuation is considered, the magnetic long range order may break down and the Lifshitz point describes a general transition from a commensurate phase to incommensurate phase.~\onlinecite{JA1,JA2,CP1,CP2}
Interestingly, in this case the position of this general Lifshitz point becomes sensitive to the magnetic anisotropy, as shown in both the analytical and numerical studies of the pure one-dimensional frustrated zigzag $XXZ$ model.~\onlinecite{JA1,JA2}
In this subsection, we consider a relatively simple case where the long range magnetic order exists on both sides of the Lifshitz point within the framework of spin wave theory.
As we have mentioned in the previous subsection, the phase boundary of different long range ordered states is deduced by comparing the energy of the states on each side.
For the sake of the completeness of discussion and the consistency of description, the magnetic anisotropy induced shift of the Lifshitz point is obtained within both the conventional and torque equilibrium schemes.

In the ferromagnetic phase, the quantum fluctuation induced corrections to the ground state energy are absent and the energy reads $E_{FM}$=$S^{2}J_{\textbf{0}}$.~\onlinecite{QM,SWT}
On the other hand, the quantum fluctuation can remarkably modify the ground state energy in the spiral phase.
In the conventional spin-wave approach, the first order corrected ground state energy in the spiral phase is
\begin{equation}
E_{S}=S(S+1)J_{\textbf{Q}_{cl}}+S\sum_{\textbf{k}}\varepsilon_{\textbf{k}}
\end{equation}
while in the torque equilibrium formulism it reads
\begin{equation}
\widetilde{E}_{S}=S(S+1)J_{\textbf{Q}}+S\sum_{\textbf{k}}\widetilde{\varepsilon}_{\textbf{k}}
\end{equation}
The main difference between the two expressions is that in the TESWT we use the quantum ordering vector $\textbf{Q}$ instead of the classical ones.
Other than that, the vacuum fluctuation energy in the torque equilibrium formulism is considered with the renormalized parameters $\widetilde{J}_{i}$.

\begin{figure*}
  \includegraphics[width=16cm]{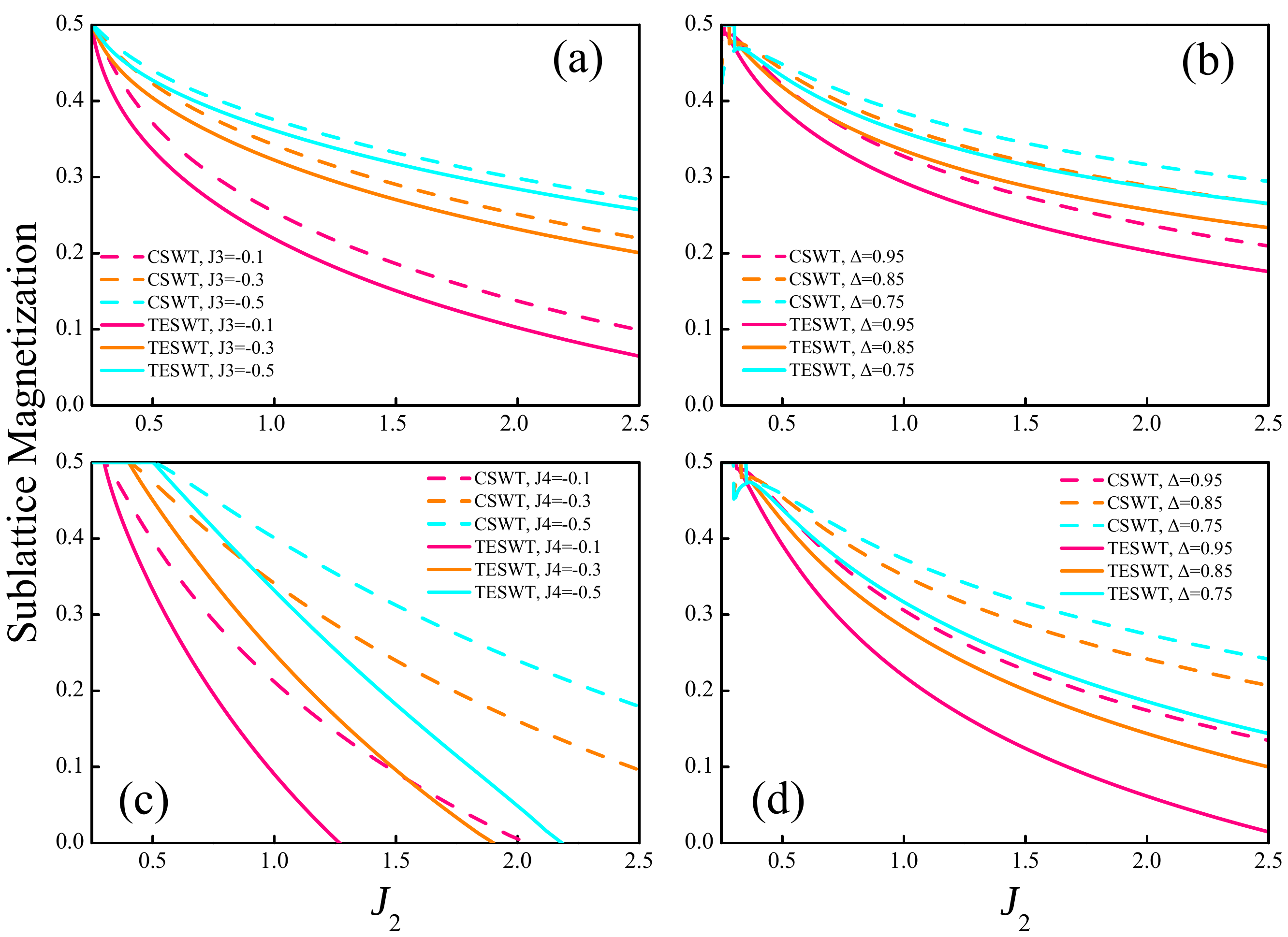}\\
  \caption{The sublattice magnetization versus $J_{2}$ in $J_{3}$- and $J_{4}$-systems with: (a) isotropic systems with various inter-chain coupling $J_{3}$; (b) anisotropic systems with various anisotropic parameter $\Delta$ and $J_{3}$ fixed to -0.1; (c) isotropic systems with various inter-chain coupling $J_{4}$; (d) anisotropic systems with various anisotropic parameter $\Delta$ and $J_{4}$ fixed to -0.1. Here the sublattice magnetization obtained through CSWT and TESWT are marked by dash and solid lines respectively. And all the exchange interactions are in units of $|J_{1}|$.}
\end{figure*}

The resultant phase diagram is shown in Fig. 4(a) and (b), corresponding to the anisotropic $J_{3}$- and $J_{4}$-system respectively.
It is clearly shown that in the classical limit, the FM/spiral phase boundary does not depend on the anisotropic energy $\Delta$.
At the same time, the magnetic anisotropy induced modification of the phase boundary is also demonstrated within the torque equilibrium spin-wave formulism: the Lifshitz point shifts towards the spiral phase as the anisotropic energy increases.
Moreover, as we've mentioned in the last subsection, the ordering vector obtained through the torque equilibrium equation can also sense the anisotropic energy induced shift of the Lifshitz point, although the phase boundary is not quantitatively the same with the formal TESWT results.
Surprisingly, however, the phase boundary is irrelevant with the $XXZ$ anisotropy within the conventional spin-wave description.
And this situation persists even when higher order perturbation to the ground state energy is considered.
Thus, the CSWT results indicate that the quantum phase boundary is exactly the same with the classical one.
As a matter of fact, the spin wave theory can only offer a qualitative rather than quantitative description of a quantum critical point.
Consequently, neither the result obtained within the conventional scheme nor that from the torque equilibrium scheme is the actual FM/spiral phase boundary.
Nevertheless, the qualitative dependence of the Lifshitz point's position on the magnetic anisotropy indicated by TESWT results should be correct.
More than that, our TESWT results are close to the finite size calculation results of the one dimensional frustrated zigzag $XXZ$ model.~\onlinecite{JA1,JA2}

\subsection{Sublattice magnetization}
Based on the quantum phase diagram that we have obtained, in this subsection we turn to the investigation of the sublattice magnetization.
The sublattice magnetization defines the validity region of the spin wave representation, which usually serves as the order parameter for general long-range ordered magnetic states.~\onlinecite{QM}
In the classical limit, the linear spin wave approximation becomes exact and the sublattice magnetization is nothing but the spin length.
The quantum fluctuation effect, on the other hand, tends to reduce the sublattice magnetization from its classical value in the long range ordered phases through the zero point fluctuation of spin waves until the breakdown of the magnetic ordered ground state.~\onlinecite{SWT}
However, in the FM state, the zero point fluctuation does not exist, and thus correspondingly the sublattice magnetization always remains its classical value at zero temperature.
As a consequence, here we only need to calculate the sublattice magnetization in the spiral phase.

In the spiral phase, as a matter of fact, an unbalanced spin Casimir torque can induce divergence in the second order correction to the sublattice magnetization in the spin wave analysis.~\onlinecite{ME}
As a result, the full one-loop calculation of the sublattice magnetization can only be performed in the torque equilibrium formulism.
However, the full one-loop calculation of the sublattice magnetization is quite tedious while the resultant second order corrected result does not show much difference from the linear approximated results as shown in Ref. 31.
Consequently, to investigate the multi-parameter dependence of the sublattice magnetization and to compare the results obtained within both CSWT and TESWT, here we only consider the linear spin wave results in both theories.
In the conventional spin-wave approach, the first order corrected sublattice magnetization is
\begin{equation}
\langle S\rangle=S\Big[1-\frac{1}{2S}\big(\sum_{\textbf{k}}\frac{A_{\textbf{k}}}{2\varepsilon_{\textbf{k}}}-1\big)\Big]
\end{equation}
while in the torque equilibrium formulism it reads
\begin{equation}
\langle \widetilde{S}\rangle=S\Big[1-\frac{1}{2S}\big(\sum_{\textbf{k}}\frac{\widetilde{A}_{\textbf{k}}}{2\widetilde{\varepsilon}_{\textbf{k}}}-1\big)\Big]
\end{equation}
The main difference between these two expressions also lies in the choice of the ordering vector and the corresponding exchange parameters.

The numerical results are shown in Fig. 5, in which the sub-plots (a) and (b) correspond to the $J_{3}$-systems and the sub-plots (c) and (d) correspond to the $J_{4}$-systems.
Different from the quantum ordering vector, the sublattice magnetization shows quite strong sensitivity to the inter-chain coupling and magnetic anisotropy.
In the ferromagnetic phase, the sublattice magnetization is simply the spin length of the system and in our case $S$=1/2.
As $J_{2}$ increases from the Lifshitz point, the sublattice magnetization of each system reduces due to strong quantum fluctuation effect enhanced by the onset of the intra-chain frustration.
This reduction is more drastic in the isotropic $J_{4}$-systems with small inter-chain coupling, e.g. in the isotropic $J_{4}$-system with $J_{4}$=-0.1, the sublattice magnetization vanishes at $J_{2}$=2 (1.25) within the CSWT (TESWT) description.
A zero sublattice magnetization represents the breakdown of the spin wave expansion around the present classical saddle point.
And in our case, this indicates that the systems may run into a spin liquid state.
Moreover, within both the conventional and torque equilibrium descriptions, there is a little kink near the Lifshitz point in the sublattice magnetization results of anisotropic $J_{3}$-and $J_{4}$-systems.
Compared with the continuum results of the isotropic systems, the appearance of the kink may indicate that the spin wave description is inadequate for a shifted Lifshitz point.
However, away from the Lifshitz point, both the CSWT and TESWT provide very reasonable results.
The suppression of the quantum fluctuation is clearly shown as the anisotropic parameter $\Delta$ and inter-chain coupling increases.
More than that, the $J_{4}$-systems always have relatively smaller sublattice magnetization compared with the $J_{3}$-systems with the same magnitude of inter-chain coupling and magnetic anisotropy.
Thus, it seems that the quantum fluctuation is stronger in the $J_{4}$-systems compared with the $J_{3}$-systems due to the crossing inter-chain coupling.

\begin{figure*}
  \includegraphics[width=18cm]{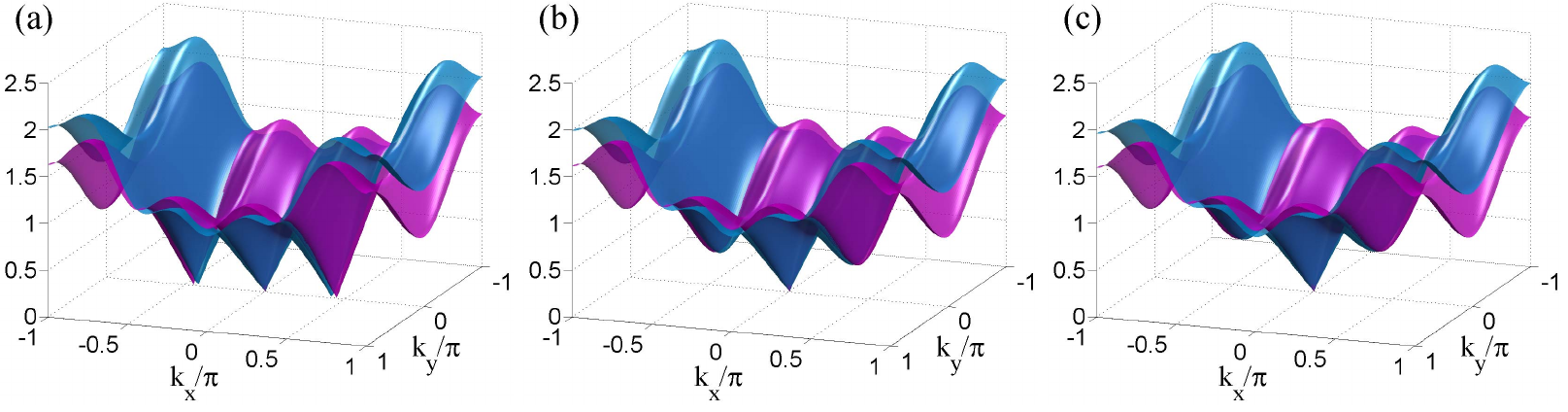}\\
  \caption{The linear spin-wave spectrum of $J_{3}$-systems, in which the blue surfaces represent the LSWT results and the purple surfaces show the TELSWT results. From left to right: (a) isotropic system with inter-chain coupling $J_{3}$=-0.3; (b) anisotropic system with inter-chain coupling $J_{3}$=-0.3 and $\Delta$=0.95; (c) anisotropic system with inter-chain coupling $J_{3}$=-0.3 and $\Delta$=0.9. Here $J_{2}$=1 and all the exchange interactions are in units of $|J_{1}|$.}
\end{figure*}

Other than these common features, the sublattice magnetization obtained in the TESWT is always smaller than that obtained through the CSWT for all the cases, which is similar to the situations in the study of anisotropic triangular AFM Heisenberg model.~\onlinecite{ME}
This fact confirms our previous estimation that the linear approximation within the TESWT is actually very close to the second order results.
Furthermore, as the quantum fluctuation effect is suppressed by increasing the inter-chain coupling or magnetic anisotropy, the sublattice magnetization obtained within the TESWT becomes closer to the ones obtained in the CSWT, which indicates that these two spin wave theories share the same classical limit.

To summarize, the sublattice magnetization obtained within the torque equilibrium linear spin wave theory (TELSWT) is qualitatively consistent with but quantitatively smaller than the ones obtained in the LSWT.
Thus the system appears to be less classical once the spin Casimir effect is taken into account.
This may shed light on our understanding of the experimental fact that some edge-shared chain cuprates usually show relatively smaller sublattice magnetization than the linear spin wave prediction.~\onlinecite{Ex1,AG3}
Together with the more accurate ordering vector predictions, it seems that our torque equilibrium approach can provide a quite good description of the quasi one-dimensional $J_{1}$-$J_{2}$ systems within the simple linear approximation.
Other than that, the TELSWT approach is much less technique-relevant compared with other sophisticated numerical and analytical methods and
it has been proven to be quite accurate for the anisotropic triangular antiferromagnets.~\onlinecite{ME}
As a consequence, the TELSWT may serve as an efficient tool for experimental fitting processes of the exchange parameters, especially for the incommensurate ordered system with strong quantum fluctuation effects.

\section{Spin-wave Spectrum}
Different from the static properties, the $1/S$ perturbative expansion results for the spin-wave spectrum have drastically qualitative difference from the results obtained by the linear approximation for noncollinear ordered antiferromagnets.
The main reason of this fact lies in the anharmonic cubic interaction terms in the effective spin wave Hamiltonian, which are forbidden in the collinear ordered states because of the unbroken $U(1)$ symmetry.~\onlinecite{Sasha2}
These anharmonic terms correspond to the coupling of the transverse and longitudinal fluctuations and can further lead to the zero temperature decays of magnon.~\onlinecite{Sasha1,Sasha2,Sasha3}
This spontaneous magnon decay effect has been theoretically proposed to occur in several magnetic systems and experimentally observed in hexagonal manganites LuMnO$_{3}$~\onlinecite{Lu} and quantum triangular antiferromagnets Ba$_{3}$CoSb$_{2}$O$_{9}$~\onlinecite{Tr} very recently.

To get into this remarkable phenomenon, the perturbative expansion has to be performed at least at one-loop order to include the self-energies contributed from the cubic vertexes.
However, the conventional spin wave expansion scheme is invalidated due to the spin Casimir effect in the incommensurate noncollinear magnetic system as explained in Ref. 31.
As a consequence, the spin wave expansion has to be performed in the torque equilibrium formulism to keep the spin wave expansion procedure away from singularities and divergences.
The dynamic properties of the system are expressed within the interacting normal magnon Green's function
\begin{equation}
G_{n}(\textbf{k},\varepsilon)=\big[\varepsilon-2S\widetilde{\varepsilon}_{\textbf{k}}-\Sigma^{tot}_{n}(\textbf{k},\varepsilon)\big]^{-1}
\end{equation}
Here $\Sigma^{tot}_{n}(\textbf{k},\varepsilon)$ represents the total normal self-energy, which in the one-loop order reads
\begin{equation}
\Sigma^{tot}_{n}(\textbf{k},\varepsilon)=\Sigma^{a}_{c}(\textbf{k})+\Sigma^{a}_{hf}(\textbf{k})+\Sigma^{a}_{3}(\textbf{k},\varepsilon)+\Sigma^{b}_{3}(\textbf{k},\varepsilon) \end{equation}
Note that $\Sigma^{a}_{c}(\textbf{k})$ only appears in the presence of the spin Casimir effect.
The poles of the normal magnon Green's function are the spin-wave spectrum, which can be obtained either by simply replacing $\varepsilon$ with linear spin-wave spectrum $2S\widetilde{\varepsilon}_{\textbf{k}}$ in the self-energies, i.e. the so-called on-sell approximation or by solving the Dyson equation self-consistently, i.e. the so-called off-shell approximation.~\onlinecite{Sasha2,Sasha3}
Both the approximation schemes can manifest the features of magnon decays but with different lineshape characteristics and associated physical interpretations, as presented below.

In the subsequent subsections, we first investigate the linear spin wave spectrum within both the CSWT and TESWT and then turn to the on-shell and off-shell calculations of the renormalized spectrum.
In order to perform a well-controlled calculation, all the numerical calculations are performed with the magnitude of the inter-chain coupling in both $J_{3}$- and $J_{4}$-systems fixed as 0.3$J_{1}$ and $J_{2}$=$|J_{1}|$.

\subsection{Harmonic approximation}
In prior to the calculation of the $1/S$ order spin-wave spectrum, it is desirable to investigate the excitation spectrum within the harmonic approximation.
In the linear spin-wave description, the magnon is a well-defined quasi-particle of the long-range ordered magnetic system with infinite long lifetime.
No spin Casimir and spontaneous magnon decay effects are involved and the spin wave spectrum is simply $2S\varepsilon_{\textbf{k}}$.
Thus, it seems that the linear spin wave spectrum has nothing to do with the nonlinear magnon decay effects.
However, the kinematic constrains for the decay features in the one-loop on-shell spin wave spectrum are actually obtained thought the linear spin wave energy equations.
In the torque equilibrium formulism, these kinematic constrains are deduced by the renormalized linear spin wave energy rather than the real harmonic one, which reads
\begin{equation}
    2S\widetilde{\varepsilon}_{\textbf{k}}=2S\sqrt{(\widetilde{J}_{\textbf{k}}-\widetilde{J}_{\textbf{Q}}+\lambda_{\textbf{Q}}+\Delta)(\widetilde{\eta}_{\textbf{k}}-\widetilde{J}_{\textbf{Q}}+\lambda_{\textbf{Q}})}
\end{equation}

\begin{figure*}
  \includegraphics[width=18cm]{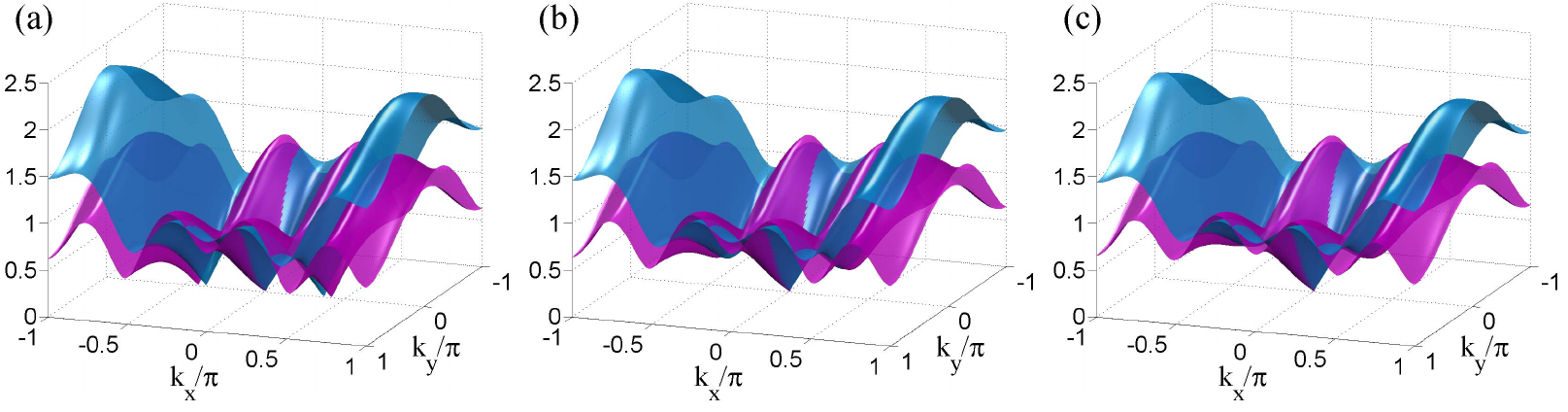}\\
  \caption{The linear spin-wave spectrum of $J_{4}$-systems, in which the blue surfaces represent the LSWT results and the purple surfaces are the TELSWT results. From left to right: (a) isotropic system with inter-chain coupling $J_{4}$=-0.3; (b) anisotropic system with inter-chain coupling $J_{4}$=-0.3 and $\Delta$=0.95; (c) anisotropic system with inter-chain coupling $J_{4}$=-0.3 and $\Delta$=0.9. Here $J_{2}$=1 and all the exchange interactions are in units of $|J_{1}|$.}
\end{figure*}

Once again, we consider the two representative $J_{3}$- and $J_{4}$-systems.
The resultant linear spin wave spectrum of the isotropic $J_{3}$-system is plotted in Fig. 6, in which the spectrum manifests obvious quasi-one dimensional characteristics: more curvy along the chain direction and less curvy along the vertical direction.
Additionally, the three Goldstone modes at $\textbf{k}$=$\textbf{0}$ and $\textbf{k}$=$\pm\textbf{Q}$ caused by the complete breaking of the $SO(3)$ rotational symmetry in the noncollinear ground state are clearly shown.
The Goldstone mode at $\textbf{k}$=$\textbf{0}$ corresponds to the in-plane sliding mode of the spiral, whereas the $\textbf{k}$=$\pm\textbf{Q}$ modes represent the out-of-plane oscillation modes related with symmetry.
As a result, these three acoustic modes have only two spin wave velocities
\begin{eqnarray}
\mathcal{V}_{\textbf{0}}&=&S\sqrt{2(J_{\textbf{0}}-J_{\textbf{Q}_{cl}})\nabla^{2}_{\textbf{k}}J_{\textbf{k}}\big|_{\textbf{k}=\textbf{Q}_{cl}}}\nonumber\\
\mathcal{V}_{\textbf{Q}}&=&S\sqrt{(J_{\textbf{0}}+J_{2\textbf{Q}_{cl}}-2J_{\textbf{Q}_{cl}})\nabla^{2}_{\textbf{k}}J_{\textbf{k}}\big|_{\textbf{k}=\textbf{Q}_{cl}}}
\end{eqnarray}
In general, these two spin wave velocities are different, i.e. one is faster than the other.
Usually, this fact actually constitutes an important decay channel from the faster Goldstone mode to the slower one, which defines the decay region as seen in such as triangular lattice antiferromagnets.~\onlinecite{Sasha1,Sasha2}
However, in the TESWT, this decay channel is actually determined by the torque equilibrium results
\begin{eqnarray}
\widetilde{\mathcal{V}}_{\textbf{0}}&=&S\sqrt{2(\widetilde{J}_{\textbf{0}}-\widetilde{J}_{\textbf{Q}})\nabla^{2}_{\textbf{k}}\widetilde{J}_{\textbf{k}}\big|_{\textbf{k}=\textbf{Q}}}\nonumber\\
\widetilde{\mathcal{V}}_{\textbf{Q}}&=&S\sqrt{(\widetilde{J}_{\textbf{0}}+\widetilde{J}_{2\textbf{Q}}-2\widetilde{J}_{\textbf{Q}})\nabla^{2}_{\textbf{k}}\widetilde{J}_{\textbf{k}}\big|_{\textbf{k}=\textbf{Q}}}
\end{eqnarray}
with the quantum ordering vector.

Once the magnetic anisotropy is considered, the out-of-plane oscillation modes at $\textbf{k}$=$\pm\textbf{Q}$ are gapped and only the in-plane sliding mode at $\textbf{k}$=$\textbf{0}$ is present.
On the other hand, the linear spin wave spectra of the corresponding $J_{4}$-systems are demonstrated in Fig. 7.
The isotropic $J_{4}$-system shows three Goldstone modes and quasi-one dimensional characteristics as well. The magnetic anisotropy manifest similar effects as in the $J_{3}$-systems.
However, there are some slight differences in the spectrum, e.g. the surface is concave or convex around some $\textbf{k}$ points.
As a matter of fact, these seemingly insignificant differences in the linear spin wave spectrum can induce very different results in terms of the spectrum at the one-loop order, as shown below.

\begin{figure*}
  \includegraphics[width=18cm]{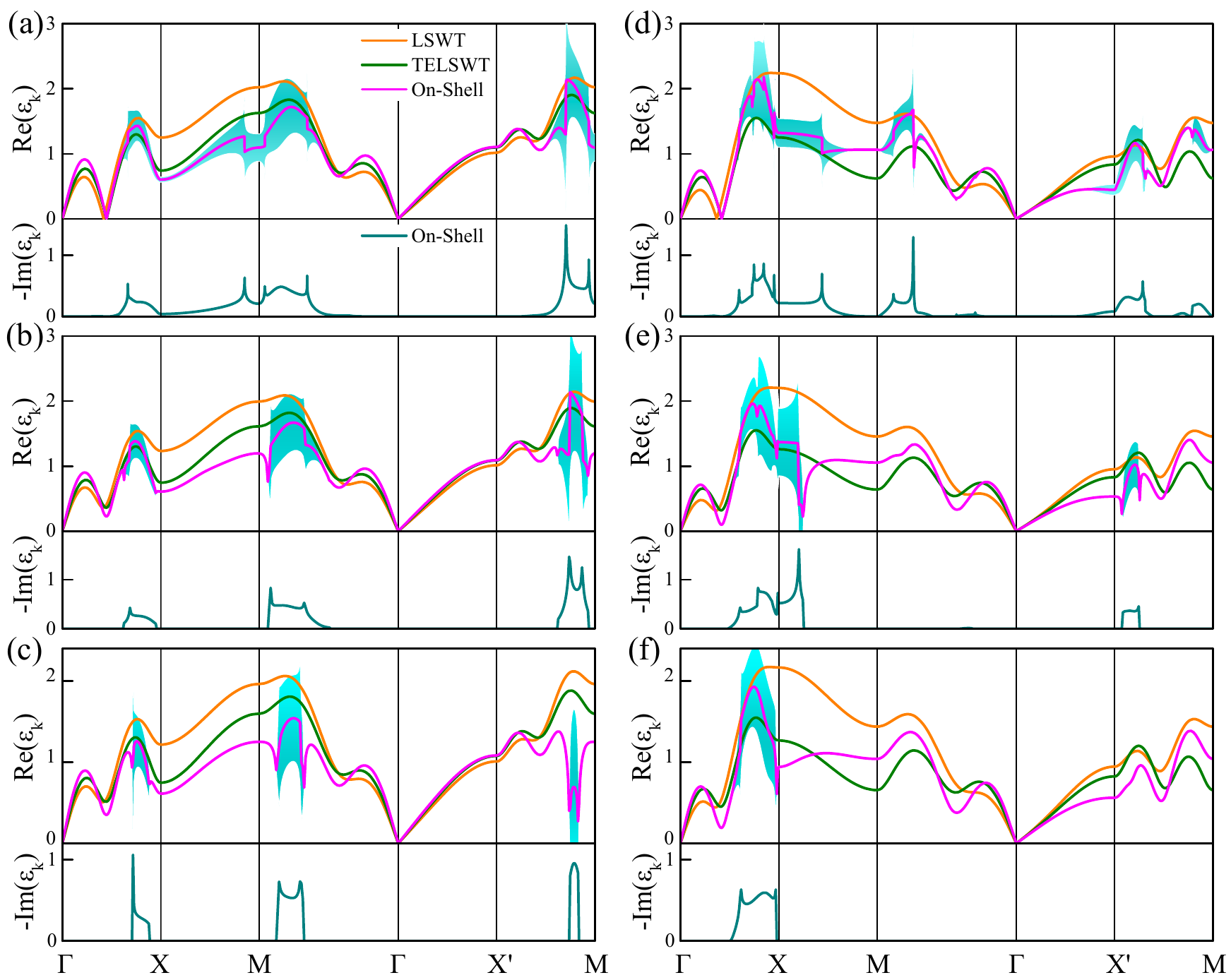}\\
  \caption{The on-shell spin wave spectrum of $J_{3}$- and $J_{4}$-systems: (a) isotropic $J_{3}$-system; (b) anisotropic $J_{3}$-system with $\Delta$=0.95; (c) anisotropic $J_{3}$-system with $\Delta$=0.9; (d) isotropic $J_{4}$-system; (e) anisotropic $J_{4}$-system with $\Delta$=0.95; (f) anisotropic $J_{4}$-system with $\Delta$=0.9. The orange lines are the results obtained in the LSWT and the green lines are the results obtained in the TELSWT. The purple and cyan lines are the real (energy) and imaginary (damping rate) parts of our on-shell $1/S$ results, respectively. And the cyan areas show the width of the spectral peaks due to the damping. Here $J_{2}$=1 and all the exchange interactions are in units of $|J_{1}|$.}
\end{figure*}

\subsection{On-shell approximation}
In the on-shell approximation, the self-energies are evaluated at the harmonic magnon energy in the torque equilibrium formulism, which represent strict $1/S$ corrections to the magnon energy.
At the one-loop order, the spin wave spectrum $\mathcal{E}_{\textbf{k}}$ obtained within the on-shell approximation can be written as
\begin{eqnarray}
   \mathcal{E}_{\textbf{k}}&=&2S\widetilde{\varepsilon}_{\textbf{k}}+\Sigma^{a}_{c}(\textbf{k})+\Sigma^{a}_{hf}(\textbf{k})\nonumber\\
                           &&+\Sigma^{a}_{3}(\textbf{k},\widetilde{\varepsilon}_{\textbf{k}})+\Sigma^{b}_{3}(\textbf{k},\widetilde{\varepsilon}_{\textbf{k}})
\end{eqnarray}
According to this expression, the one-loop spin wave spectrum $\mathcal{E}_{\textbf{k}}$ can be easily obtained by numerical integration of the self-energies.

The numerical results for the $J_{3}$- and $J_{4}$-systems are demonstrated in Fig. 8 along some representative symmetry directions in the Brillouin zone (BZ).
The spin wave spectrum show prominent features of the magnon damping in the major part of the BZ with giant imaginary part of $\mathcal{E}_{\textbf{k}}$.
As we have mentioned before, this remarkable magnon damping originates from the coupling between the single-particle excitations and the two-particle continuum determined by the anharmonic cubic terms.
A feature of the results is that the magnon damping and the renormalization of the spin wave spectrum are stronger at large momenta.
Additionally, there are many substantial singularities in both the real and imaginary parts of $\mathcal{E}_{\textbf{k}}$, which manifest themselves in the form of jump-like discontinuities and spike-like peaks.~\onlinecite{Sasha1,Sasha2,Sasha3}
The origin of these singularities is due to the intersection of the single-magnon branch with the line of the van Hove saddle point singularities in the two-magnon continuum.
The analytical properties of these singularities have deep connection with the integration dimensionality of the self-energies, which have been profoundly discussed in Ref. 25.
In two-dimensional systems, they fulfill the Kramers-Kronig relations between the real and imaginary parts of the one-loop on-shell spin wave energy as:
\begin{equation}
\mathrm{Re}(\mathcal{E}_{\textbf{k}})\simeq \mathrm{sgn}(\delta\textbf{k}),~~~~\mathrm{Im}(\mathcal{E}_{\textbf{k}})\simeq- \mathrm{ln}\Big(\frac{\Lambda}{|\delta\textbf{k}|}\Big)
\end{equation}
or the other way around:
\begin{equation}
\mathrm{Re}(\mathcal{E}_{\textbf{k}})\simeq \mathrm{ln}\Big(\frac{\Lambda}{|\delta\textbf{k}|}\Big),~~~~\mathrm{Im}(\mathcal{E}_{\textbf{k}})\simeq- \Theta(\delta\textbf{k})
\end{equation}
Here $\delta\textbf{k}$=$\textbf{k}$-$\textbf{k}^{*}$ with $\textbf{k}^{*}$ represents the location of the singularities in $\textbf{k}$ space, $\Lambda$ is the cut-off parameter determined by characteristic size of the singular region, $\mathrm{sgn}(x)$ stands for the sign function and $\Theta(x)$ is the Heaviside step function.
When the inter-plane coupling is considered, the self-energies integration becomes three dimensional and the associated logarithmic peaks become the square root ones.

\begin{figure*}
  \includegraphics[width=18cm]{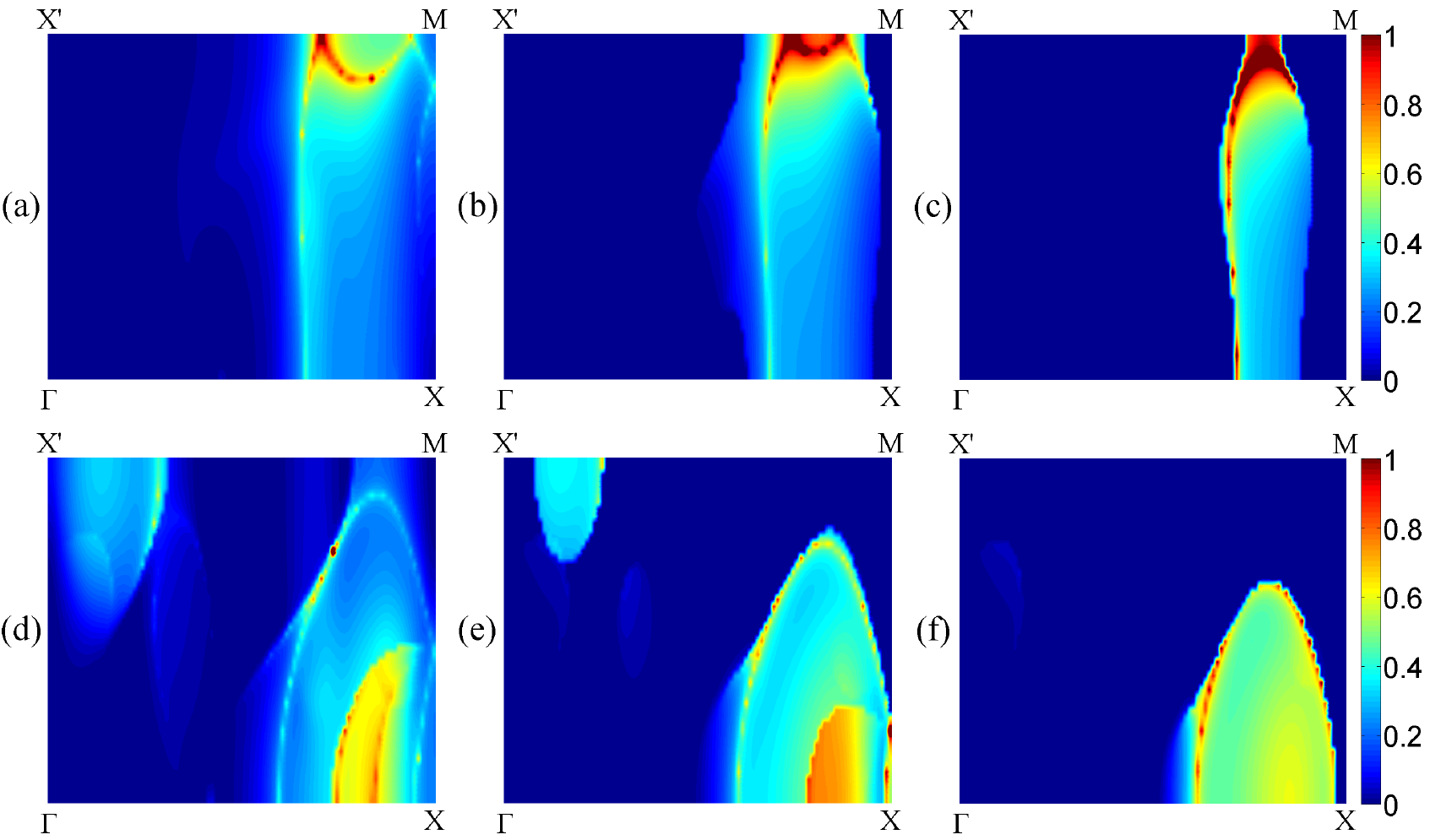}\\
  \caption{The intensity plots of the on-shell magnon decay rate of $J_{3}$- and $J_{4}$-systems: (a) isotropic $J_{3}$-system; (b) anisotropic $J_{3}$-system with $\Delta$=0.95; (c) anisotropic $J_{3}$-system with $\Delta$=0.9; (d) isotropic $J_{4}$-system; (e) anisotropic $J_{4}$-system with $\Delta$=0.95; (f) anisotropic $J_{4}$-system with $\Delta$=0.9. Here all the exchange interactions are in units of $|J_{1}|$.}
\end{figure*}

It is worth noting that the imaginary part of the spin wave energies is only non-vanishing in some region of the BZ, i.e. the so-called magnon decay region.
The threshold boundary of this region is determined by the kinematic constrains that follow the momentum and energy conservation in the two-particle decay process
\begin{equation}
\widetilde{\varepsilon}_{\textbf{k}}=\widetilde{\varepsilon}_{\textbf{p}}+\widetilde{\varepsilon}_{\textbf{k}-\textbf{p}}
\end{equation}
and the extremum condition of the two-particle continuum, or equivalently the spin wave velocity equation
\begin{equation}
\widetilde{\mathcal{V}}_{\textbf{p}}=\widetilde{\mathcal{V}}_{\textbf{k}-\textbf{p}}
\end{equation}
These equations are expressed simply based on the harmonic approximation for the magnon energies, which can only determine the decay boundary at the one-loop order.
As the anharmonic terms induced renormalization effect is considered, the kinematic constrains are modified as well.
As a consequence, the threshold boundary changes as the higher order spin wave processes are taken into account.
However, the decay boundary obtained within the harmonic approximation is usually considered to be immensely instructive.
One of the reasons is the higher order renormalized decay region usually shows little difference from the harmonic one, as demonstrated in triangular lattice antiferromagnetic Heisenberg (TLAH) model.~\onlinecite{Sasha1,Sasha2}
Another reason, may be one of the most important reasons, is that a consideration of the threshold boundary within the harmonic approximation can be carried out analytically.
More than that, not only the decay region but also the location and decay channel of the singularities can be obtained.

In spite of the methodological convenience and the analytical availability, the decay boundary analysis can be difficult to carry out for higher order or in a self-consistent manner if not impossible.
Additionally, the decay boundary defines the region where magnon decay is allowed, but the spin wave excitations are not strongly damped within the whole region inside the decay boundary.
However, only the considerable broad peaks caused by the strong magnon decay effect can be experimentally observed because of the nonzero temperature and the finite resolution of the detector.
Together with the fact that in our case, the decay region actually covers most of the area in the BZ, we then turn to the analysis of the strong damping region instead of the harmonic boundary.
In order to take into account the damping rate of the spin wave excitations, it is convenient to introduce the magnon decay rate, which can be expressed as
\begin{equation}
\mathcal{R}_{\textbf{k}}=-\mathrm{Im}\big[\Sigma^{tot}_{n}(\textbf{k},\varepsilon)\big]
\end{equation}
Within the one-loop on-shell approximation, it can be written as~\onlinecite{Sasha1,Sasha2,Sasha3}
\begin{equation}
\mathcal{R}_{\textbf{k}}=\frac{\pi}{2}\sum_{\textbf{p}}\widetilde{\Gamma}^{2}_{1}(\textbf{p};\textbf{k})
\cdot\delta(\widetilde{\varepsilon}_{\textbf{k}}-\widetilde{\varepsilon}_{\textbf{p}}-\widetilde{\varepsilon}_{\textbf{k}-\textbf{p}})
\end{equation}
which is nothing but the imaginary part of the self-energy $\Sigma^{a}_{3}(\textbf{k},\widetilde{\varepsilon}_{\textbf{k}})$ and manifests similar form to the Fermi's golden-rule expression.

The numerical results of the spin wave decay rate in one quarter of the BZ are shown in Fig. 9, in which the strong decay region is intuitively demonstrated.
It is straightforward to verify from the decay boundary analysis that the bright line and sharp boundaries shown in Fig. 9 belong to some specific decay channels. However, not all the decay boundaries obtained from the harmonic approximation analysis have an obvious demonstration in the decay rate intensity plots.
In addition, the demonstrated decay pattern of each system shows very good consistency with the corresponding on-shell spectrum.
Surprisingly, the decay pattern and corresponding on-shell spin wave spectrum for isotropic $J_{3}$- and $J_{4}$-systems show drastically different features, which can also be read out from the on-shell spectrum.
The isotropic $J_{4}$-system shows far more singularities than the $J_{3}$-system, e.g. there are four singularities along the $\Gamma$-X direction in the $J_{4}$-system while only one along the same path in the $J_{3}$-system.

Although it appears interesting, the pattern of the singularity distribution may not be experimentally observable because the singularities can be smoothed, considering the higher order contributions.
Nevertheless, there are several remarkable differences that may survive even in the self-consistent calculation.
First, the strong damping region is around the M point in the $J_{3}$-system while in the $J_{4}$-system the magnon decay is stronger around the X point.
Next, there is a wide decay region around the X$'$ point in $J_{4}$-system while no decay at all near the same region in the $J_{3}$-system.
This is a direct consequence of the difference in the harmonic spin wave spectrum between the $J_{3}$- and $J_{4}$-systems.
The last but not the least, another striking feature is the nearly flat mode along the X-M direction of the renormalized spin wave spectrum for the $J_{4}$-system, which is very different from the results of the corresponding $J_{3}$-system and the classical $J_{4}$-system.
Upon a close examination, we find that this flat mode is induced by the self-energy $\Sigma^{a}_{c}(\textbf{k})$ that is contributed from the spin Casimir effect.
The conventional self-energies that describe the three- and four-magnon interactions are usually negative, which correspond to the downward renormalization of the excitation spectrum in the usual magnon decay cases.
However, the self-energy $\Sigma^{a}_{c}(\textbf{k})$ has a huge positive contribution to the total self-energy around the M point in the $J_{4}$-system.
As a consequence, the renormalization of the excitation spectrum turns upward compared with the TELSWT results, and then the flat mode along the X-M direction appears.
Note that the total renormalization of the spin wave spectrum is still downward compared with the LSWT results, and thus the qualitative magnon decay arguments remain reliable.~\onlinecite{Sasha1,Sasha2,Sasha3}

\begin{figure*}
  \includegraphics[width=18cm]{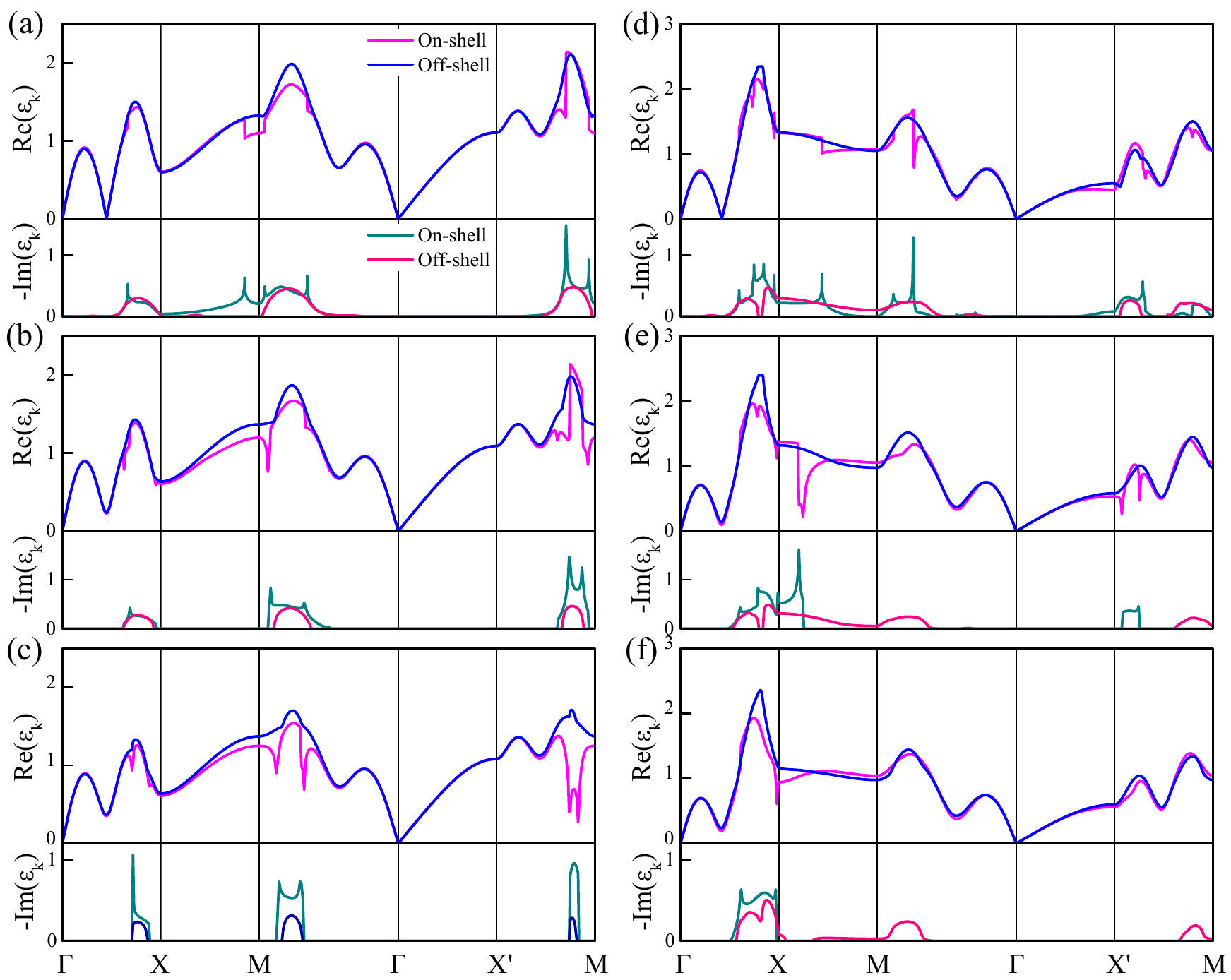}\\
  \caption{The off-shell spin wave spectrum of $J_{3}$- and $J_{4}$-systems: (a) isotropic $J_{3}$-system; (b) anisotropic $J_{3}$-system with $\Delta$=0.95; (c) anisotropic $J_{3}$-system with $\Delta$=0.9; (d) isotropic $J_{4}$-system; (e) anisotropic $J_{4}$-system with $\Delta$=0.95; (f) anisotropic $J_{4}$-system with $\Delta$=0.9. The blue and pink lines are the real (energy) and imaginary (damping rate) parts of our off-shell $1/S$ results, respectively. And the on-shell results are also plotted as a comparison. Here all the exchange interactions are in units of $|J_{1}|$. }
\end{figure*}

As the magnetic anisotropy is introduced, the on-shell spin wave spectra of the $J_{3}$- and $J_{4}$-systems are both drastically modified.
Similar to the classical case, the Goldstone modes at $\textbf{k}$=$\pm\textbf{Q}$ are gapped and only the Goldstone mode at $\textbf{k}$=$\textbf{0}$ is present.
However, the quantum fluctuation effect can modify the classical spectrum and the spin wave gap at $\textbf{k}$=$\pm\textbf{Q}$ in the $J_{3}$- and $J_{4}$-systems are both drastically downward renormalized.
Furthermore, the $J_{4}$-system appears to be less sensitive to the magnetic anisotropy for the far smaller spin wave gap compared with the $J_{3}$-system with the same anisotropy energy.
This fact may shed light on our understanding of the experimental results on LiCuVO$_{4}$, in which the inelastic neutron scattering data show nearly zero spin wave gap at $\textbf{k}$=$\pm\textbf{Q}$, while the electron spin resonance results indicate a 6$\%$ magnetic anisotropy in the system.~\onlinecite{Ex1,ESR}
Additionally, the spontaneous magnon decay region is drastically reduced due to the suppression of the quantum fluctuation effect and the kinematic condition for decays caused by the magnetic anisotropy.
Surprisingly, in spite of the sharp reduction of the decay region, the decay rate is less reduced and the prominent differences in the decay pattern between the $J_{3}$- and $J_{4}$-systems still exist.
At the same time, the flat mode in the isotropic $J_{4}$-system also survives in the anisotropic cases.
As a consequence, the spontaneous magnon decay effects can be expected to be robust in the quasi-one dimensional $J_{1}$-$J_{2}$ magnetic systems and the decay pattern manifests different features for different types of inter-chain couplings.

\begin{figure*}
  \includegraphics[width=18cm]{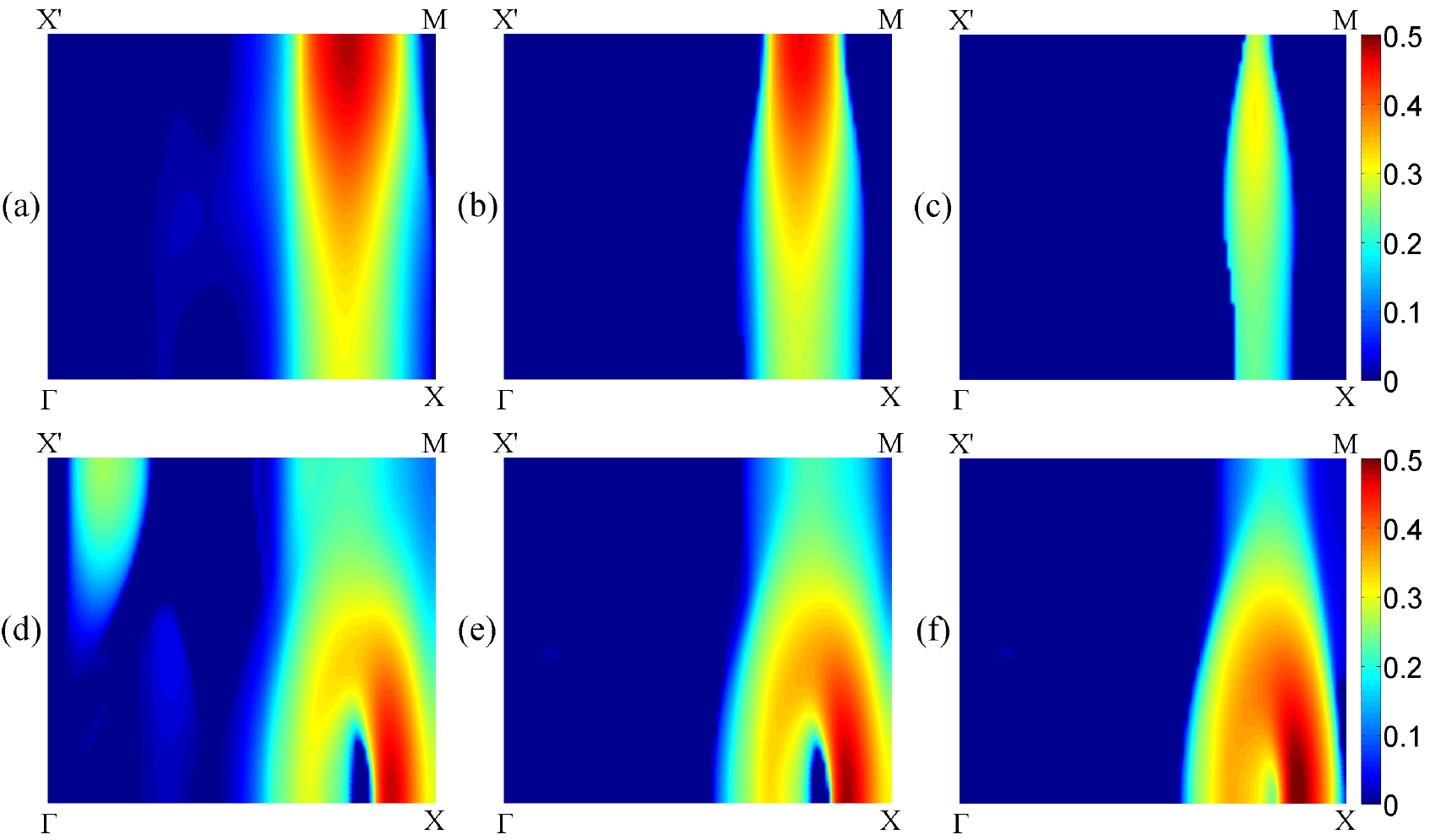}\\
  \caption{The intensity plots of the off-shell magnon decay rate of $J_{3}$- and $J_{4}$-systems: (a) isotropic $J_{3}$-system; (b) anisotropic $J_{3}$-system with $\Delta$=0.95; (c) anisotropic $J_{3}$-system with $\Delta$=0.9; (d) isotropic $J_{4}$-system; (e) anisotropic $J_{4}$-system with $\Delta$=0.95; (f) anisotropic $J_{4}$-system with $\Delta$=0.9. Here all the exchange interactions are in units of $|J_{1}|$.}
\end{figure*}

\subsection{Off-shell approximation}
As a matter of fact, the unusual singularities in the on-shell spin wave spectrum actually signify a breakdown of the standard spin wave expansion.
Consequently, a self-consistent calculation has to be performed in order to obtain the actual dynamic properties of the system.~\onlinecite{Sasha2,Sasha4,Sasha5,Sasha6}
In this subsection, we turn to one of the self-consistent schemes: the off-shell approximation. Within this approximation, the self-energies are evaluated in a self-consistent way by allowing the finite lifetime of the magnon at the very beginning, while the magnons created during the decay process remain stable.

The spin wave spectrum $\mathcal{D}_{\textbf{k}}$ within the off-shell approximation can be obtained by self-consistently solving the Dyson equation
\begin{eqnarray}
   \mathcal{D}_{\textbf{k}}&=&2S\widetilde{\varepsilon}_{\textbf{k}}+\Sigma^{a}_{c}(\textbf{k})+\Sigma^{a}_{hf}(\textbf{k})\nonumber\\
   &&+\Sigma^{a}_{3}(\textbf{k},\mathcal{D}^{*}_{\textbf{k}})+\Sigma^{b}_{3}(\textbf{k},\mathcal{D}^{*}_{\textbf{k}})
\end{eqnarray}
where $\mathcal{D}^{*}_{\textbf{k}}$ is the complex conjugate of $\mathcal{D}_{\textbf{k}}$ following the methodological discussion in Ref. 25 on the proper sign of the imaginary part of the decay-like self-energy.
Rewriting the above equation explicitly for the real and imaginary parts, the original Dyson equation becomes the following equation sets:
\begin{eqnarray}
\mathrm{Re}(\mathcal{D}_{\textbf{k}})&=&2S\widetilde{\varepsilon}_{\textbf{k}}+\Sigma^{a}_{c}(\textbf{k})+\Sigma^{a}_{hf}(\textbf{k})\nonumber\\
                             &&+\mathrm{Re}\big[\Sigma^{a}_{3}(\textbf{k},\mathcal{D}^{*}_{\textbf{k}})+\Sigma^{b}_{3}(\textbf{k},\mathcal{D}^{*}_{\textbf{k}})\big] \nonumber\\
\mathrm{Im}(\mathcal{D}_{\textbf{k}})&=&\mathrm{Im}\big[\Sigma^{a}_{3}(\textbf{k},\mathcal{D}^{*}_{\textbf{k}})+\Sigma^{b}_{3}(\textbf{k},\mathcal{D}^{*}_{\textbf{k}})\big]
\end{eqnarray}
According to this expression, the off-shell one-loop spin wave spectrum $\mathcal{D}_{\textbf{k}}$ can be easily obtained by numerically solving the integration equations.
At the same time, the magnon decay rate within the off-shell approximation can be simply expressed as
\begin{equation}
\mathcal{R}_{\textbf{k}}=-\mathrm{Im}(\mathcal{D}_{\textbf{k}})
\end{equation}
from which the self-consistent decay pattern can be directly obtained.

The off-shell spin wave spectrum for the $J_{3}$- and $J_{4}$-systems are demonstrated in Fig. 10 along the same representative symmetry directions with the on-shell results.
As a self-consistent method, the spin wave spectrum and decay rate obtained within the off-shell approach contains the contributions beyond the one-loop order.
Consequently, the remarkable singularities in the on-shell one-loop spectrum are regularized within the off-shell scheme.
Additionally, the spin wave spectra obtained within the off-shell approximation are stretched upwards a little compared with the on-shell ones, which may be caused by the over-estimation of the energy shifts in the on-shell scheme.
In spite of this upward renormalization, the off-shell spin wave gap at $\textbf{k}$=$\pm\textbf{Q}$ in the $J_{4}$-systems remains smaller compared with the $J_{3}$-systems with the same anisotropy energy as in the on-shell cases.
On the other hand, the nearly flat mode along the X-M direction in $J_{4}$-systems in less significant than that in the on-shell cases but remains quite different with the $J_{3}$- and classical $J_{4}$-systems, and thus may still serve as a characteristic of the quantum $J_{4}$-systems.

The corresponding spin wave decay rate data in one quarter of the BZ are shown in Fig. 11.
While the jump-like discontinuities and spike-like peaks disappear in both the real and imaginary parts of the spin wave energy, the magnon decay rate remains significant throughout BZ.~\onlinecite{Sasha2,Sasha3}
The overall shape of the decay region is very different from the on-shell cases, the reason of which may lie in the sensitivity of the quasi-one dimensional systems.
Nevertheless, the characteristic differences in the decay pattern between the $J_{3}$- and $J_{4}$-systems that we discussed in the on-shell case survive in the off-shell results.
The strong damping region still lies around the M and X points in the $J_{3}$- and $J_{4}$-systems respectively and the wide decay region around the X$'$ point in the isotropic $J_{4}$-system still exists while the same region in the isotropic $J_{3}$-system remains no decay at all.
Moreover, the effect of the magnetic anisotropy on the decay rate is further enhanced within the off-shell scheme.
In particular, both the off-shell approximated decay region and decay rate are dramatically reduced compared with the corresponding on-shell results in the $J_{3}$-systems.
On the other hand, the $J_{4}$-systems show less sensitivity on the area of the decay region, although the magnitude of the decay rate is much smaller than the on-shell predictions.

Different from the case in the TLAH model,~\onlinecite{Sasha1,Sasha2} the off-shell decay region is slightly different from the on-shell predictions, which may be caused by the extraordinary sensitivity of the quasi-one dimensional frustrated $J_{1}$-$J_{2}$ systems.
Other than that, a more astonishing feature of the off-shell results is the appearance of non-decay area in the center of the strong decay region around the X point in the isotropic and weakly anisotropic $J_{4}$-systems.
In this sudden non-decay region, the imaginary part of the off-shell spectrum vanishes and the real component manifests a flat-top peak.
Although the off-shell decay pattern is not necessarily identical with the on-shell prediction, the on-shell broadening in this sudden non-decay area reaches nearly one-half of the spectrum.
Consequently, one may expect that it is unlikely that this spectrum broadening can disappear in the self-consistent calculation, especially in the area isolated and surrounded by strong decay region.
Moreover, this sudden non-decay region disappears once the magnetic anisotropy parameter exceeds some critical value $\Delta_{c}$ and the Dyson equation finds the solution with finite broadening as expected.

To verify this strange characteristic and further investigate the magnon decay dynamics with the off-shell approximation, we introduce the poles function $\mathcal{P}_{\textbf{k}}$, which is the inverse of the normal magnon Green's function and expressed within the one-loop approximation as
\begin{eqnarray}
   \mathcal{P}_{\textbf{k}}&=&\varepsilon-2S\widetilde{\varepsilon}_{\textbf{k}}-\Sigma^{a}_{c}(\textbf{k})-\Sigma^{a}_{hf}(\textbf{k})\nonumber\\
   &&-\Sigma^{a}_{3}(\textbf{k},\varepsilon)-\Sigma^{b}_{3}(\textbf{k},\varepsilon)
\end{eqnarray}
The zeros of this function represent the poles of the normal magnon Green's function, which are equivalent with the solutions of the Dyson equation.
In order to analyze the sudden non-decay area, the self-energies are obtained with frequency $\varepsilon$ scans through the real axis.
This scheme is very similar to the one adopted in the calculation of the spectral function, which is discussed in details in the next subsection.
In order to perform a concrete calculation, we choose the representative point $\textbf{k}$=$(0.82\pi,0,0)$ which lies right in the middle of the sudden non-decay region.
Additionally, we calculate another representative point $\textbf{k}$=$(\pi,0,0)$, which locates near the sudden non-decay area and lies in the decay region in the isotropic case.
More than that, as a comparison, we also calculate the point $\textbf{k}$=$(0,0.2\pi,0)$, which locates far way from both the decay and the sudden non-decay region.
Furthermore, to clarify the related spin wave decay dynamics, we introduce the two-magnon density of states (DOS)~\onlinecite{Sasha2}
\begin{eqnarray}
   \mathcal{M}_{\textbf{k}}&=&\sum_{\textbf{p}}\delta(\varepsilon-\widetilde{\varepsilon}_{\textbf{p}}-\widetilde{\varepsilon}_{\textbf{k}-\textbf{p}})
\end{eqnarray}
which has van Hove singularities that can cross the single particle spectrum, therefore can be responsible for magnon decays.

\begin{figure}
  \includegraphics[width=8.6cm]{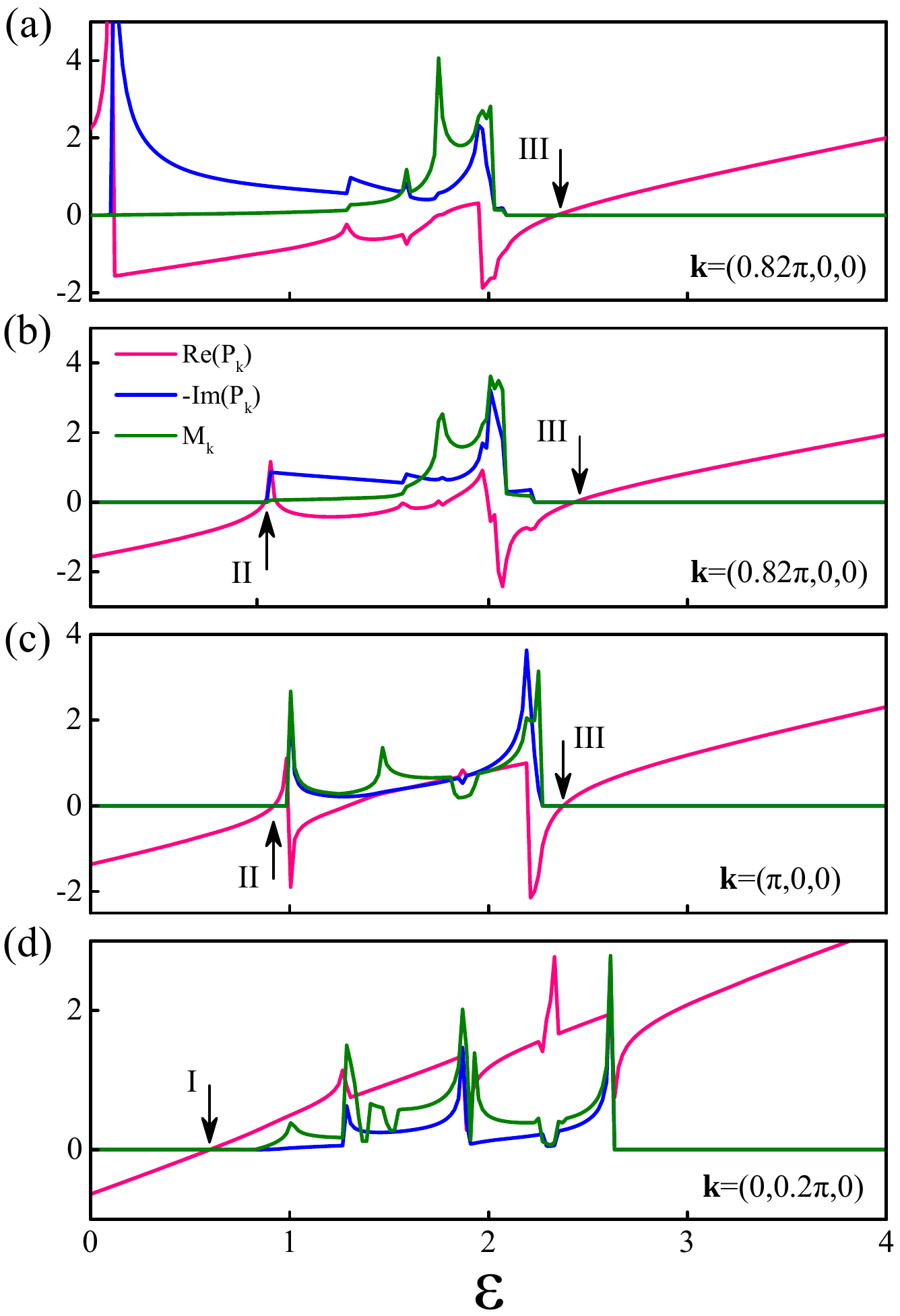}\\
  \caption{The poles function $\mathcal{P}_{\textbf{k}}$ together with the two-magnon DOS $\mathcal{M}_{\textbf{k}}$ of $J_{4}$-systems: (a) isotropic system for $\textbf{k}$=$(0.82\pi,0,0)$; (b) anisotropic system with $\Delta$=0.9 for $\textbf{k}$=$(0.82\pi,0,0)$; (c) isotropic system for $\textbf{k}$=$(\pi,0,0)$; (d) anisotropic system with $\Delta$=0.9 for $\textbf{k}$=$(0,0.2\pi,0)$. The blue and purple lines are the real and imaginary parts of the poles function and the green lines are the two-magnon DOS, respectively. Three types of crossing points are ladled with I, II and III (see main text).}
\end{figure}

The results of the calculation of various functions at different representative $\textbf{k}$ points are shown in Fig. 12.
The demonstrated correspondence between the step-like and spike-like singularities in $\mathrm{Re}(\mathcal{P}_{\textbf{k}})$ and $\mathrm{Im}(\mathcal{P}_{\textbf{k}})$ is very similar to that in the on-shell cases, which is also the direct consequence of the Kramers-Kronig relations.
While the correspondence between the $\mathcal{P}_{\textbf{k}}$ and $\mathcal{M}_{\textbf{k}}$ is more delicate, corresponding to certain magnon decay channels, we are interested in the non-decay solutions of the Dyson equation. Here we fucus on a specific kind of crossing points between $\mathrm{Re}(\mathcal{P}_{\textbf{k}})$ and the real $\varepsilon$ axis with $\mathrm{Im}(\mathcal{P}_{\textbf{k}})$=0.
Surprisingly though, such crossing point is not unique in most cases that we have investigated and there are three types.

The first type of such crossing points are the most conventional ones, which represent the true well-defined quasi-particles with zero damping and in our case only appear in Fig. 12(d).~\onlinecite{SWT}
This type of crossing points locate well beneath the two-magnon DOS, and thus only exist in the real non-decay region.
The second type of such crossing points are more delicate and have deep connection with the singular behavior near the bottom of the two-magnon DOS.
Different with the first type, this type of solutions only appear in the decay region.
Strictly speaking, this type of crossing points should not be considered as a well-defined quasi-particle state because they are caused by the singular behavior of the one-loop self-energies, thus may disappear when the one-loop singularities are regularized by the higher order approximation.
However, as a matter of fact, they have very obvious demonstration in the one-loop spectral function results and are referred to as "edge" singularities in the study of the TLAH model.~\onlinecite{Sasha2}
Contrary to the previous two types, the last type of crossing points locate well beyond the two-magnon DOS, which are actually the single-particle state pushed out of the two-magnon continuum and referred to as "antibonding" magnon states.~\onlinecite{Sasha6}
These states exist in almost all the cases that we have investigated.
Although in the usual cases only the "bonding" magnon states are considered as the right solutions of the Dyson function, the "antibonding" states also exist in some special cases but normally do not mixed up with the "bonding" ones because of the large energy gap between them.
However, this conventional picture breaks down in our case of the sudden non-decay area.

Note the continues transition between the sudden non-decay region and the strong decay region shown in Fig. 10(d) and Fig. 11(d), it means that the "bonding" states are actually moving towards the upward located "antibonding" states when approaching the sudden non-decay area.
As a consequence, the strange sudden non-decay area observed in the off-shell spectrum is not a true non-decay region but a region in which the "bonding" and "antibonding" magnon states are degenerate.
And the flat-top peak is a direct demonstration of the "antibonding" spectrum, which is rather flat around the X point.
Although this strange phenomenon appears to be interesting, the degeneration between these single-particle states is usually not robust, i.e. their appearance depend on the approximation one used in the calculation.
Thus the whole sudden non-decay region may be simply an accidental product of our one-loop Dyson scheme, which can disappear once another self-consistent scheme is adopted.
At last, we would like to mention that all these three types of solutions can be clearly demonstrated in the spectral function obtained within the one-loop approximation, which can be considered as another evidence of our analysis.

\begin{figure*}
  \includegraphics[width=18cm]{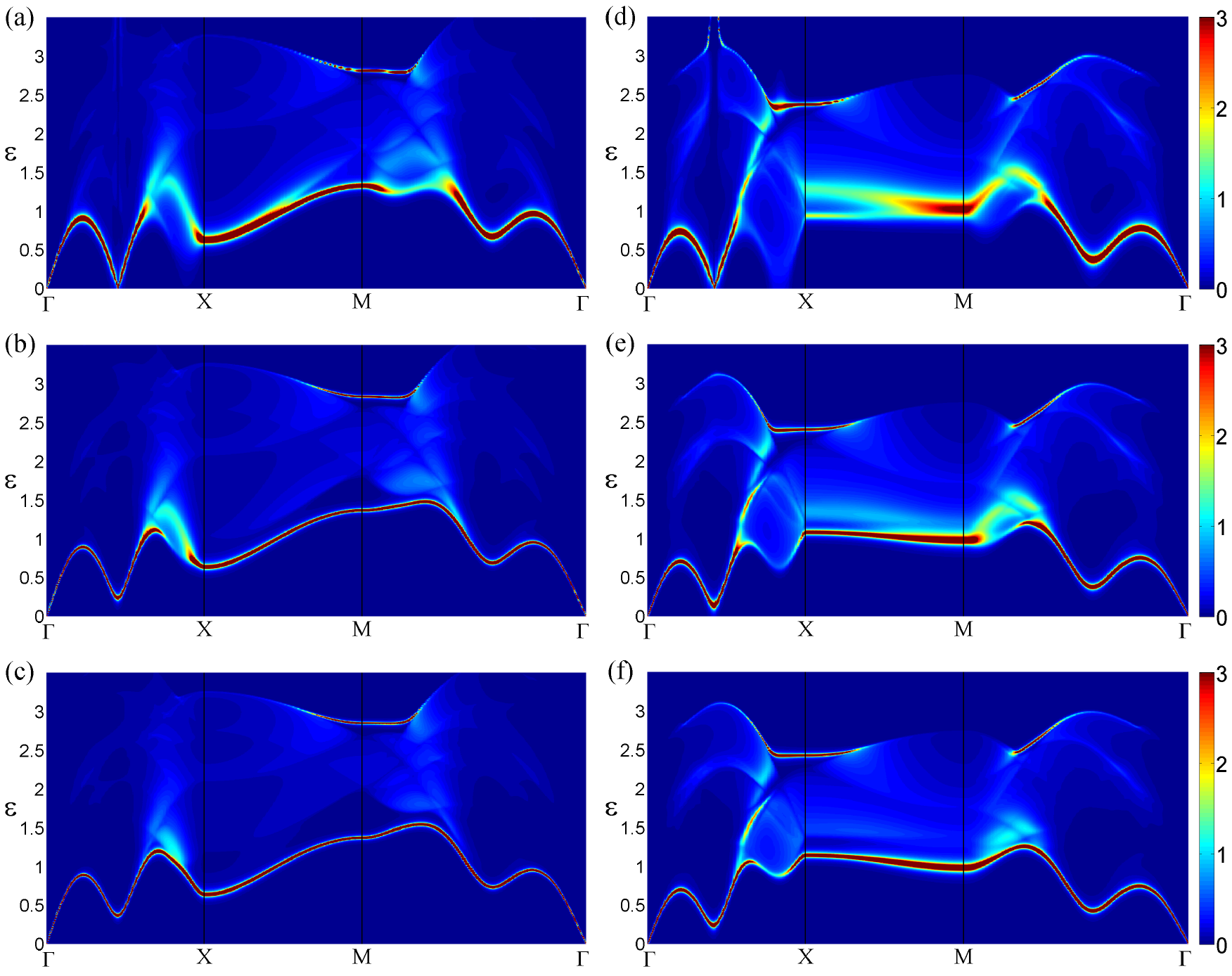}\\
  \caption{The intensity plots of the spectral function $\mathcal{A}(\textbf{k},\varepsilon)$ of $J_{3}$- and $J_{4}$-systems: (a) isotropic $J_{3}$-system; (b) anisotropic $J_{3}$-system with $\Delta$=0.95; (c) anisotropic $J_{3}$-system with $\Delta$=0.9; (d) isotropic $J_{4}$-system; (e) anisotropic $J_{4}$-system with $\Delta$=0.95; (f) anisotropic $J_{4}$-system with $\Delta$=0.9. Here all the exchange interactions are in units of $|J_{1}|$.}
\end{figure*}

\section{Spectral function}
In order to further investigate the actual dynamic properties of the system we turn to the spectral function, which contain a detailed information on the spin wave energy renormalization and the magnon decay rate.
The diagonal component of the spectral function is defined by the imaginary part of the normal magnon Green's function through the expression~\onlinecite{Sasha5,Sasha6,Mahan}
\begin{equation}
\mathcal{A}(\textbf{k},\varepsilon)=-\frac{1}{\pi}\mathrm{Im}\big[G_{n}(\textbf{k},\varepsilon)\big]
\end{equation}
Within the leading one-loop approximation the normal magnon Green's function is given by Eq. (36).
The spectral function $\mathcal{A}(\textbf{k},\varepsilon)$ is deeply connected with the dynamical structure factor $\mathcal{S}(\textbf{k},\varepsilon)$ which is directly measured in inelastic neutron scattering experiments.~\onlinecite{Sasha5}
Generally, $\mathcal{S}(\textbf{k},\varepsilon)$ also contains contributions from the off-diagonal and two-particle correlations, which we leave to the next section for a detailed discussion.
In this subsection, we focus on the spectral function of the system, which actually provides the major component of the dynamical structure factor but much easier to analyze.

In the classical limit, the spectral function is a $\delta$ function located at $2S\varepsilon_{\textbf{k}}$ for any momentum $\textbf{k}$ in the BZ.~\onlinecite{Mahan}
Similar to the spin wave spectrum, the spectral function is strongly renormalized by the quantum fluctuation as well.~\onlinecite{Sasha2,Sasha4,Sasha5,Sasha6}
In the absence of the intrinsic damping, the spin wave energy is real and the quasiparticle peak in $\mathcal{A}(\textbf{k},\varepsilon)$ occurs precisely at $\mathcal{D}_{\textbf{k}}$, the off-shell spin wave spectrum obtained by solving the Dyson equation.
However, when the spontaneous magnon decay occurs the spin wave energy acquires a significant imaginary part.
As a consequence, the location and broaden width of the quasiparticle peak in the spectral function is different with the solution of the Dyson equation because $\mathcal{A}(\textbf{k},\varepsilon)$ is only defined on the real $\varepsilon$ axis.
In additional to the single-particle peaks, the spectral function is expected to exhibit an incoherent component, which represents the contribution from the two-particle continuum due to the nonorthogonality between the single- and two-particle excitations.
In the calculation of the spectral function, the normal magnon Green's function is obtained with frequency scans through all the possible energies.
Consequently, the consideration of the spectral function is also beyond the $1/S$ expansion, which is analog with the off-shell approximation and the poles function that we've discussed.
On the other hand, the numerical integration of the self-energies is performed with an artificial broadening of $\sigma$=0.01.
And the intensity plots of the spectral function for the $J_{3}$- and $J_{4}$- systems are demonstrated in Fig. 13.

A feature of the demonstrated spectral function observed in the whole BZ is the downward renormalization of the single-particle dispersion, which is consistent with our previous on-shell and off-shell results.
In addition, the spontaneous magnon decay induced broadening of the spectrum can be directly observed around certain $\textbf{k}$ points in the BZ.
Other than that, the characteristics differences that we expected still remain in the spectral function.
As shown clearly in the results of the isotropic $J_{3}$- and $J_{4}$-systems, the strong damping region still lies around the M and X point respectively.
However, this feature disappears in the anisotropic $J_{3}$-system with $\Delta$=0.95, in which the spectrum broadening is more remarkable around the X point and nearly indiscernible around the M point.
This may be a consequence of the fact that the decay pattern manifested in the spectral function is different from the on-shell and off-shell results, which further indicates that the quasi-one dimensional $J_{1}$-$J_{2}$ system is very sensitive.
More than that, the $J_{3}$-systems can still be concluded to be more sensitive to the magnetic anisotropy than the $J_{4}$-systems.
On the other hand, the nearly flat mode along the X-M direction in $J_{4}$-systems in more significant than that in the on-shell and off-shell cases due to the remarkable broadening of the spectrum.
And this salient broadening of the flat mode shrinks dramatically once the magnetic anisotropy is introduced.

Another prominent feature is the appearance of "pseudo"-quasiparticle peaks that we've discussed in the previous subsection.
One interesting type of "pseudo"-quasiparticle peaks are the so called "edge" singularities, which usually locate near the bottom of the spectral function.~\onlinecite{Sasha2}
They only exist in the decay region and manifest as a bright "edge" of the spectral function.
More interestingly, the "edge" in the spectral function of the isotropic $J_{4}$-system goes to zero around the X point, which is consistent with the poles function results.
This strange behavior may not be deeply connected with the sudden non-decay area issue because the former case own wider region than the latter one.
As the magnetic anisotropy is introduced and the magnon decay get suppressed, the "edge" singularities move towards the true quasi-particle peaks and finally merge into them when the magnon decay is absent.

The most astonishing and misleading type of "pseudo"-quasiparticle peaks are the "antibonding" magnon states lie around the top of the spectral function.
As a matter of fact, they are actually true single-magnon state at least within the one-loop approximation as we've shown in the previous subsection.
However, this type of excitations only exist in certain region of the BZ and the energy scale of them are much larger than the usual spin wave excitation that we're interested in, thus normally they are not considered as one part of the excitation spectrum.~\onlinecite{Sasha6}
But this is not the case for the isotropic $J_{4}$-system, in which the "bonding" and "antibonding" single-magnon states are degenerate near X point as shown in Fig. 13(d).
This degeneration exactly explain the appearance of the sudden non-decay area obtained within the off-shell approximation.
As clearly demonstrated in the spectral function, this region is actually a strongly decay region rather than the non-decay area as the off-shell results indicated. At last, we would like to stress that this degeneration and even the "antibonding" magnon states themselves can be just an accidental product of our one-loop approximation.
And to our best of knowledge, there is not yet a conclusive evidence about the existence of the "antibonding" magnon states in any real system.
Nevertheless, they still exist within the one-loop scheme and can lead to very obvious phenomena in the associated off-shell excitation spectrum and the spectral function results, thus need to be clarified at some level.

\section{Dynamic Structure Factor}
\begin{figure*}
  \includegraphics[width=18cm]{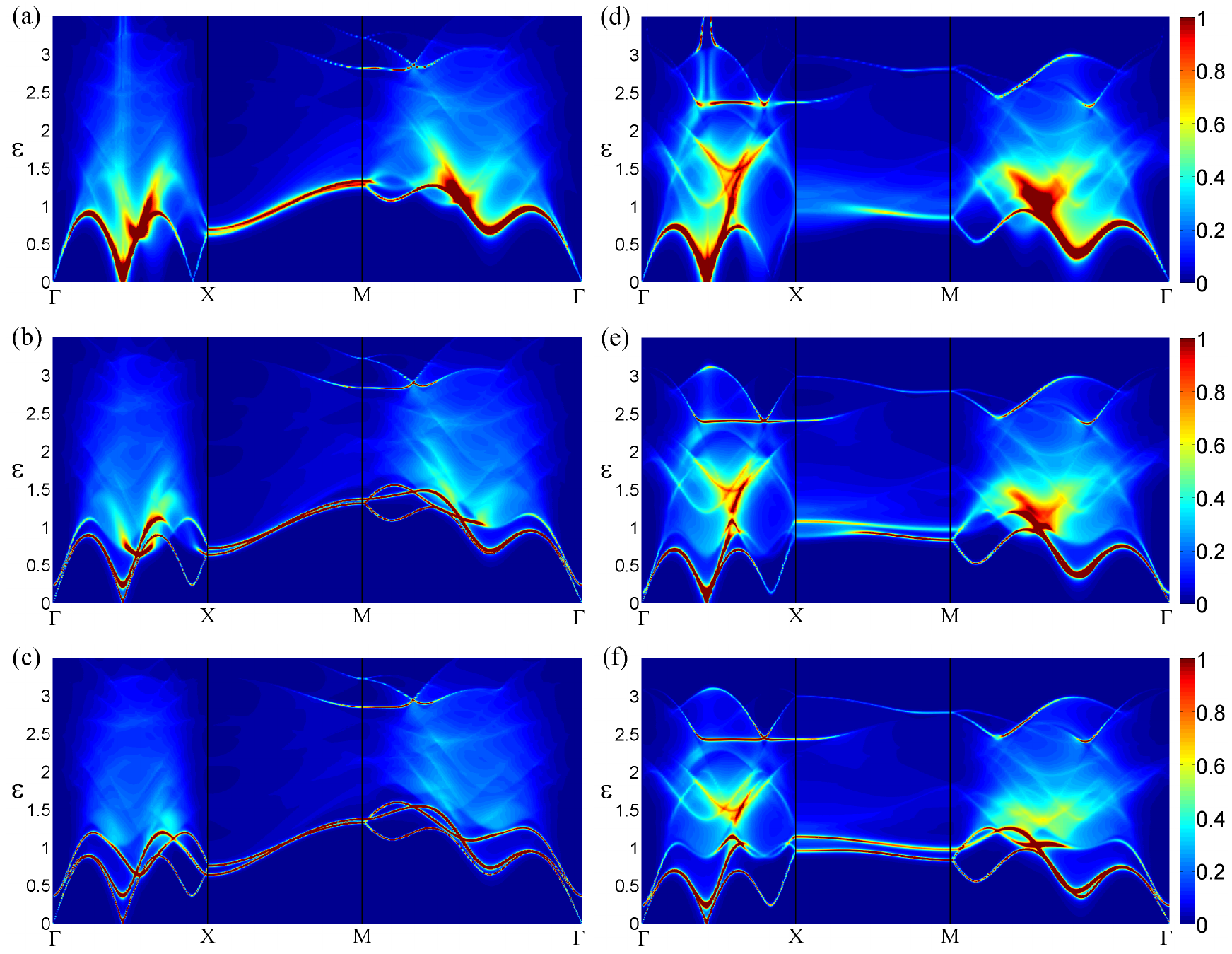}\\
  \caption{The intensity plots of the dynamic structure factor $\mathcal{S}(\textbf{k},\varepsilon)$ of $J_{3}$- and $J_{4}$-systems: (a) isotropic $J_{3}$-system; (b) anisotropic $J_{3}$-system with $\Delta$=0.95; (c) anisotropic $J_{3}$-system with $\Delta$=0.9; (d) isotropic $J_{4}$-system; (e) anisotropic $J_{4}$-system with $\Delta$=0.95; (f) anisotropic $J_{4}$-system with $\Delta$=0.9. Here all the exchange interactions are in units of $|J_{1}|$.}
\end{figure*}
The consideration of the dynamic structure factor $\mathcal{S}(\textbf{k},\varepsilon)$ in noncollinear antiferromagnets has been given for cases of the square-lattice antiferromagnets in the magnetic field, space isotropic triangular lattice antiferromagnets and more recently the anisotropic models on kagome lattice.
However, the noncollinear magnetic order in all these systems is commensurate and not plagued with the divergences and singularities caused by the spin Casimir effect.~\onlinecite{ME}
For the more general incommensurate cases, the spin wave expansion has to be performed within the so called torque equilibrium approach.
While the corresponding formulism of $\mathcal{S}(\textbf{k},\varepsilon)$ has not yet been established within the TESWT.
In this section, we investigate the dynamic structure factor of the $J_{3}$- and $J_{4}$-systems within the torque equilibrium scheme.
One of our goals of the present section is to generalize the calculation of $\mathcal{S}(\textbf{k},\varepsilon)$ to the TESWT.
Our formulism is similar to the isotropic triangular lattice case,~\onlinecite{Sasha6} but with modifications due to the torque equilibrium condition, thus we revisit the whole derivation for comparison and completeness.
And our second goal is to provide the first explicit theoretical results for the dynamic structure factor of the quasi-one dimensional $J_{1}$-$J_{2}$ models to guide experimental inelastic neutron scattering measurements in realistic materials.

The dynamic structure factor is nothing but the spin-spin correlation function which can be directly probed in inelastic neutron scattering experiments.
By definition, it can be expressed as
\begin{equation}
   \mathcal{S}^{\mu\nu}(\textbf{k},\varepsilon)=\int^{\infty}_{-\infty}\frac{dt}{2\pi}e^{i\varepsilon t}\big\langle S^{\mu}_{\textbf{k}}(t)S^{\nu}_{-\textbf{k}}(0)\big\rangle
\end{equation}
with $\mu,\nu\in(a,b,c)$ and $S^{\mu}_{\textbf{k}}(t)$ represents the Fourier transformed spin operator.
Usually, it is convenient to define the dynamical correlation functions as
\begin{equation}
   \mathcal{G}^{\mu\nu}(\textbf{k},\varepsilon)=-i\int^{\infty}_{-\infty}\frac{dt}{2\pi}e^{i\varepsilon t}\big\langle \mathcal{T}S^{\mu}_{\textbf{k}}(t)S^{\nu}_{-\textbf{k}}(0)\big\rangle
\end{equation}
with $\mathcal{T}$ represents the time-ordered operator.
As a consequence, the dynamic structure factor is then connected with dynamical correlation function through the fluctuation-dissipation theorem~\onlinecite{Mahan}
\begin{equation}
   \mathcal{S}^{\mu\nu}(\textbf{k},\varepsilon)=-\frac{1}{\pi}\mathrm{Im}\big[\mathcal{G}^{\mu\nu}(\textbf{k},\varepsilon)\big]
\end{equation}

The inelastic neutron-scattering cross section is actually a linear combination of the diagonal components of the spin-spin correlation function with momentum dependent prefacers according to the experimental settings.~\onlinecite{NS}
For simplicity, however, we do not assume a particular experimental geometry in this section and only consider the total structure factor
\begin{equation}
   \mathcal{S}^{tot}(\textbf{k},\varepsilon)=\mathcal{S}^{aa}(\textbf{k},\varepsilon)+\mathcal{S}^{bb}(\textbf{k},\varepsilon)+\mathcal{S}^{cc}(\textbf{k},\varepsilon)
\end{equation}
Rewriting each component in the twisted frame we obtain
\begin{eqnarray}
\mathcal{S}^{aa}(\textbf{k},\varepsilon)&=&\frac{1}{4}\Big[\mathcal{S}^{xx}(\textbf{k}+\textbf{Q},\varepsilon)+\mathcal{S}^{xx}(\textbf{k}-\textbf{Q},\varepsilon)\nonumber\\
                                         &&+\mathcal{S}^{zz}(\textbf{k}+\textbf{Q},\varepsilon)+\mathcal{S}^{zz}(\textbf{k}-\textbf{Q},\varepsilon)\Big]  \nonumber\\
                                         &&+\frac{i}{4}\Big[\mathcal{S}^{zx}(\textbf{k}+\textbf{Q},\varepsilon)-\mathcal{S}^{zx}(\textbf{k}-\textbf{Q},\varepsilon) \nonumber\\
                                         &&-\mathcal{S}^{xz}(\textbf{k}+\textbf{Q},\varepsilon)+\mathcal{S}^{xz}(\textbf{k}-\textbf{Q},\varepsilon)\Big]\nonumber\\
\mathcal{S}^{bb}(\textbf{k},\varepsilon)&=&\mathcal{S}^{yy}(\textbf{k},\varepsilon),~~~~\mathcal{S}^{cc}(\textbf{k},\varepsilon)=\mathcal{S}^{aa}(\textbf{k},\varepsilon)
\end{eqnarray}
Once again, we use the quantum ordering vector instead of the classical one.
The total structure factor can be divided in terms of the diagonal and mixed contributions as~\onlinecite{Sasha6,LT}
\begin{equation}
\mathcal{S}^{tot}(\textbf{k},\varepsilon)=\mathcal{S}^{diag}(\textbf{k},\varepsilon)+\mathcal{S}^{mix}(\textbf{k},\varepsilon)
\end{equation}
and the diagonal one can be further separated into transverse and longitudinal parts as
\begin{equation}
\mathcal{S}^{diag}(\textbf{k},\varepsilon)=\mathcal{S}^{L}(\textbf{k},\varepsilon)+\mathcal{S}^{T}(\textbf{k},\varepsilon)
\end{equation}
with
\begin{eqnarray}
\mathcal{S}^{L}(\textbf{k},\varepsilon)&=&\frac{1}{2}\Big[\mathcal{S}^{zz}(\textbf{k}+\textbf{Q},\varepsilon)+\mathcal{S}^{zz}(\textbf{k}-\textbf{Q},\varepsilon)\Big]  \nonumber\\
\mathcal{S}^{T}(\textbf{k},\varepsilon)&=&\frac{1}{2}\Big[\mathcal{S}^{xx}(\textbf{k}+\textbf{Q},\varepsilon)+\mathcal{S}^{xx}(\textbf{k}-\textbf{Q},\varepsilon)\Big] \nonumber\\
&&+\mathcal{S}^{yy}(\textbf{k},\varepsilon) \nonumber\\
\mathcal{S}^{mix}(\textbf{k},\varepsilon)&=&\frac{i}{2}\Big[\mathcal{S}^{zx}(\textbf{k}+\textbf{Q},\varepsilon)-\mathcal{S}^{zx}(\textbf{k}-\textbf{Q},\varepsilon) \nonumber\\
                                         &&-\mathcal{S}^{xz}(\textbf{k}+\textbf{Q},\varepsilon)+\mathcal{S}^{xz}(\textbf{k}-\textbf{Q},\varepsilon)\Big]
\end{eqnarray}

On the other hand, the relation between the dynamical correlation functions and the magnon Green's functions can be obtained by
transforming the spin operators into the Holstein-Primakoff representation~\onlinecite{HP} and proceeding with the Bogoliubov transformation.
In this rather standard procedure, the Hartree-Fock approximation has been made as we did for the quartic interaction terms and
the final results about the transverse components read
\begin{eqnarray}
\mathcal{G}^{xx}(\textbf{k},\varepsilon)&=&\frac{S}{2}\mathcal{C}_{x}^{2}(\widetilde{u}_{\textbf{k}}+\widetilde{v}_{\textbf{k}})^{2}
           \big[G_{n}(\textbf{k},\varepsilon)+G_{n}(-\textbf{k},-\varepsilon)\nonumber\\
           &&+2G_{a}(\textbf{k},\varepsilon)\big] \nonumber\\
\mathcal{G}^{yy}(\textbf{k},\varepsilon)&=&\frac{S}{2}\mathcal{C}_{y}^{2}(\widetilde{u}_{\textbf{k}}-\widetilde{v}_{\textbf{k}})^{2}
           \big[G_{n}(\textbf{k},\varepsilon)+G_{n}(-\textbf{k},-\varepsilon)\nonumber\\
           &&-2G_{a}(\textbf{k},\varepsilon)\big]
\end{eqnarray}
with
\begin{eqnarray}
\mathcal{C}_{x}&=&1-\frac{1}{4S}\sum\limits_{\textbf{p}}(2\widetilde{v}^{2}_{\textbf{p}}+\widetilde{u}_{\textbf{p}}\widetilde{v}_{\textbf{p}}) \nonumber\\
\mathcal{C}_{y}&=&1-\frac{1}{4S}\sum\limits_{\textbf{p}}(2\widetilde{v}^{2}_{\textbf{p}}-\widetilde{u}_{\textbf{p}}\widetilde{v}_{\textbf{p}})
\end{eqnarray}
where $G_{a}(\textbf{k},\varepsilon)$ is the anomalous magnon Green's functions which can be written as
\begin{equation}
G_{a}(\textbf{k},\varepsilon)=G_{n}(\textbf{k},\varepsilon)\Sigma^{tot}_{a}(\textbf{k},\varepsilon)G_{n}(-\textbf{k},-\varepsilon)
\end{equation}
Here $\Sigma^{tot}_{a}(\textbf{k},\varepsilon)$ represents the total anomalous self-energy, which in one-loop order can be expressed explicitly as
\begin{equation}
\Sigma^{tot}_{a}(\textbf{k},\varepsilon)=\Sigma^{b}_{c}(\textbf{k})+\Sigma^{b}_{hf}(\textbf{k})+\Sigma^{c}_{3}(\textbf{k},\varepsilon)+\Sigma^{d}_{3}(\textbf{k},\varepsilon)
\end{equation}
Note that the sign of the infinitely small imaginary part in the self-energies $\Sigma^{a}_{3}(\textbf{k},\varepsilon)$ and $\Sigma^{c}_{3}(\textbf{k},\varepsilon)$ have to be changed to account for the retarded self-energies and ensure the correct odd-frequency dependence of the imaginary part of the magnetic suspendibility as stressed in in the study of the TLAH model.~\onlinecite{Sasha6}

Similar to the calculation of the excitation spectrum and spectral function, we restrict ourselves to the leading $1/S$ order of the spin wave theory, thus several simplifications can be made following Ref. 46.
Firstly, the anomalous magnon Green's functions is of the next order in $1/S$ classification compared to the normal ones $G_{n}(\textbf{k},\varepsilon)$ due to Eq. (64), therefore can be neglected.
Secondly, the $G_{n}(-\textbf{k},-\varepsilon)$ term in Eq. (62) is off-resonance compared to $G_{n}(\textbf{k},\varepsilon)$ and contains no poles for $\varepsilon$$>$0, thus has no contribution to the structure factor.
All together, the approximated expressions of the transverse dynamical structure factor read
\begin{eqnarray}
   \mathcal{S}^{xx}(\textbf{k},\varepsilon)&=&\frac{S}{2}\mathcal{C}_{x}^{2}(\widetilde{u}_{\textbf{k}}+\widetilde{v}_{\textbf{k}})^{2}\mathcal{A}(\textbf{k},\varepsilon)\nonumber\\
   \mathcal{S}^{yy}(\textbf{k},\varepsilon)&=&\frac{S}{2}\mathcal{C}_{y}^{2}(\widetilde{u}_{\textbf{k}}-\widetilde{v}_{\textbf{k}})^{2}\mathcal{A}(\textbf{k},\varepsilon)
\end{eqnarray}
which are simply linear combination of the diagonal component of the spectral functions $\mathcal{A}(\textbf{k},\varepsilon)$ and $\mathcal{A}(\textbf{k}\pm\textbf{Q},\varepsilon)$.

The consideration of the longitudinal component is more delicate than the transverse ones, which is determined by the correlation function
\begin{equation}
   \mathcal{S}^{zz}(\textbf{k},t)=\langle \delta S^{z}_{\textbf{k}}(t)\delta S^{z}_{-\textbf{k}}(0)\big\rangle
\end{equation}
with
\begin{equation}
   \delta S^{z}_{\textbf{k}}=-\sum\limits_{\textbf{p}}a^{\dagger}_{\textbf{p}}a_{\textbf{k}+\textbf{p}}
\end{equation}
As a consequence, it is in fact an $1/S$ order smaller than the transverse correlation functions.
Therefore, just linear spin wave results are adequate to account for the $1/S$ order contributions.
And in the torque equilibrium formulism, the result becomes
\begin{eqnarray}
   \mathcal{S}^{zz}(\textbf{k},\varepsilon)&=&\frac{1}{2}\sum\limits_{\textbf{p}}\delta(\varepsilon-\widetilde{\varepsilon}_{\textbf{p}}-\widetilde{\varepsilon}_{\textbf{k}-\textbf{p}})       \cdot(\widetilde{u}_{\textbf{p}}\widetilde{v}_{\textbf{k}-\textbf{p}}\nonumber\\
                                           &&+\widetilde{v}_{\textbf{p}}\widetilde{u}_{\textbf{k}-\textbf{p}})^{2}
\end{eqnarray}
The applicability of these approximation have been examined at length in Ref. 46 for the TLAH model, in which more details can be found.
At the same time, we also ignore the contributions from the mixed part of the structure factor and assume that $\mathcal{S}^{tot}(\textbf{k},\varepsilon)\approx\mathcal{S}^{diag}(\textbf{k},\varepsilon)$.
And the numerical results of the total dynamical structure factor for the $J_{3}$-and $J_{4}$-systems are demonstrated in Fig. 14, which are obtained based on Eq. (66) and (69).

The dynamical structure factor for the $J_{3}$-and $J_{4}$-systems are plotted along the same representative symmetry paths with the spectral function results.
The complicated view of the plot is the consequence of the superposition of three $\textbf{k}$-modulated $\mathcal{A}(\textbf{k},\varepsilon)$ terms and a background of two $\mathcal{S}^{zz}(\textbf{k},\varepsilon)$ terms with the incommensurate ordering vector $\textbf{Q}$.
Similar with the spectral function results, the dynamical structure factor also shows sharp single-particle peaks as well as the substantial two-particle continuum contributions.
In addition, the spontaneous magnon decay induced broadening of the quasi-particle peaks and the high energy "antibonding" single magnon states are manifested clearly.
Although the strong decay region is not easy to read from such a complicated demonstration, the nearly flat mode in the $J_{4}$-systems is clearly shown.
Altogether, the total structure factor exhibits a complex but consistent landscape of magnetic excitations with our previous spectrum and spectral function results.

\section{Discussion and Conclusion}
There are two main motivations for the work presented in this paper.
The first one is to extend and comprehend our previously developed spin wave expansion approach to more realistic multi-parameter cases, where we can test the applicability of our theory.
To accomplish this, we propose a so-called one-parameter renormalization approximation scheme, which can be considered as a good approximation for the quasi-one dimensional systems and its applicability can be further verified by the ordering vector results.
Based on this scheme, a series of magnetic properties can be obtained within the TESWT and no nonphysical singularities and divergences appear in the presence of the spin Casimir effect.
Not only the $1/S$ expansion results, but also the linear approximated torque equilibrium approach can indicate some important information such as the QObD effect and the magnetic anisotropy induced modification of the FM/spiral Lifshitz point.
Thus, generally speaking, it seems that the TESWT can efficiently describe the magnetic dynamics in the incommensurate noncollinear long-range ordered magnetic systems where the CSWT may break down, at least qualitatively.
Moreover, the TESWT can take into account part of the quantum fluctuation effects even at the linear approximation level, thus may serve as a more reliable parameter fitting tool than conventional LSWT.
The reliability of the TESWT as a parameter fitting tool has been discussed in detail for Cs$_{2}$CuCl$_{4}$ in our previous work.~\onlinecite{ME}
However, for systems concerned in this article, a quantitative investigation of this issues is still impracticable due to the lacking of exact measurements of the exchange parameters.

Our second motivation is to perform a systematic investigation of the spin wave dynamics of the long-range ordered quasi-one dimensional $J_{1}$-$J_{2}$ system.
This system has attracted much interest recently due to the high field spin multipolar phases observed in edge-shared chain cuprates.~\onlinecite{Ex3,Ex4,Ex5}
However, the associated spin wave expansion studies of the related zero field spiral state was lacking due to the divergent problems caused by spin Casimir effect.
In this work, we perform the calculation of the excitation spectrum in both the on-shell and off-shell approximation scheme.
The highlights of the anomalous features of the spectrum that should be observable in experiments are the substantial broadening of quasi-particle peaks due to the spontaneous magnon decay and strong deviations from the LSWT results, which usually serve as foundation of parameter fitting processes in experiments.
Moreover, the dependence of the spin wave spectrum on the inter-chain coupling and magnetic anisotropy is uncovered, which may serve as a prob of different types of coupling.
These remarkable distinct features are qualitative different decay pattern, different sensitivity to the magnetic anisotropy and the appearance of the nearly flat mode.
And to highlight these differences and compare the on-shell and off-shell results in the whole BZ, we investigate the magnon decay rate in both on- and off-shell scheme.

In additional to these physical features, there are also some methodology related problems in the spectrum calculation such as the "sudden non-decay region" in the off-shell results.
To clarify these issues, we introduce the poles function and two-magnon density of states and perform a detailed calculation at $\textbf{k}$ points in and outside this region.
And it turns out that the "sudden non-decay region" is actually a decay region but with degenerate "bonding" and "antibonding" single-magnon states, which is consistent with the on-shell predictions.
Other than that, we further investigate the spectral function as verification to the spectrum results, in which the degeneration of "bonding" and "antibonding" single-magnon states are directly demonstrated.
Furthermore, we develop the analytical theory of the dynamical structure factor within the torque equilibrium formulism and present the explicit numerical results, which complete our previous analysis and can be considered as guide for the inelastic neutron scattering experiments.

The present analysis can be straightforwardly generalized to the cases where the frustrated $J_{1}$-$J_{2}$ chains are coupled in a three dimensional manner.
The three dimensional coupling can have plentiful rich and varied types, all of which can suppress the quantum fluctuation effect and change the dimension of integration in the calculation of the self-energies.
As a consequence, some of our results can be modified accordingly, such as the spikelike logarithmic peaks appear in the on-shell spectrum.
However, we expect that the main spontaneous magnon decay related effects and the sensitivity of the magnetic dynamics to different types of inter-chain coupling can still survive due to the quasi-one dimensionality, frustrated intra-chain coupling and associated incommensurate long-range spiral order.
Moreover, our investigation can also be directly applied to the large $S$ cases, where the spin wave prediction is more reliable but the over all damping will be smaller as a result of the suppression of the quantum fluctuation due to the large spin length.
On the other hand, the long-range spiral order can be more robust in the compounds with large spin length.
In addition, as the measurements of the lifetimes of spin expiation in the inelastic neutron and resonant inelastic x-ray scattering experiments are expected to improve dramatically in the future,~\onlinecite{NS,RIXS} this will allow for the detailed analysis of the spontaneous magnon decays.

To summarize, we have presented a systematic spin wave analysis of quasi-one dimensional $J_{1}$-$J_{2}$ systems with different types of inter-chain couplings and $XXZ$ anisotropy including various static and dynamical magnetic properties.
Our results provide the first determination of the full magnetic properties for the long-range ordered quasi-one dimensional $J_{1}$-$J_{2}$ system within the framework of TESWT.
In addition, these detailed calculations provide a direct analytical scheme to investigate the spin dynamics for the realistic materials as well as a guide for
experimental observation of the quasi-one dimensional spontaneous magnon decay effects and inter-chain coupling dependent dynamic features.
Thus this work presents the full landscape of the nonlinear spin wave dynamics in the quasi-one dimensional FM-AFM frustrated $J_{1}$-$J_{2}$ systems.

\section{Acknowledgment}
The authors are indebted to A.L.Chernyshev for illuminating discussions.
This work was supported by the National Natural Science Foundation of China (Grants No. 11234005 and No. 51431006), China Postdoctoral Science Foundation (2015M571729) and the National 973 Projects of China (Grant No. 2015CB654602).




\begin{thebibliography}{99}

\bibitem{SL} L. Balents, Nature (London) {\bf 464}, 199 (2010).

\bibitem{Wen} X.-G. Wen, \emph{Quantum Field Theory of Many-Body Systems} (Oxford University Press, New York, 2004).

\bibitem{OR} O. A. Starykh, Rep. Prog. Phys. {\bf 78}, 052502 (2015).

\bibitem{QM} \emph{Quantum Magnetism}, edited by U. Schollwock, J. Richter, D. Farnell, and R. Bishop (Springer, Berlin, 2004).

\bibitem{SN1} A. V. Chubukov, \prb {\bf 44}, 4693 (1991).

\bibitem{J1} A. A. Nersesyan, A. O. Gogolin, and F. H. L. Essler, \prl {\bf 81}, 910 (1998).

\bibitem{JH1} D. V. Dmitriev, and V. Y. Krivnov, \prb {\bf 73}, 024402 (2006).

\bibitem{JH2} T. Hikihara, L. Kecke, T. Momoi, and A. Furusaki, \prb {\bf 78}, 144404 (2008).

\bibitem{JH3} J. Sudan, A. L\"{u}scher, and A. M. L\"{a}uchli, \prb {\bf 80}, 140402 (2009).

\bibitem{JA1} D. V. Dmitriev, and V. Y. Krivnov, \prb {\bf 77}, 024401 (2008).

\bibitem{JA2} S. Furukawa, M. Sato, and S. Onoda, \prl {\bf 105}, 257205 (2010).

\bibitem{Sk1} J. Sirker, \prb {\bf 81}, 014419 (2010).

\bibitem{Sk2} J. Ren, and J. Sirker, \prb {\bf 85}, 140410 (2012).

\bibitem{Ex1} M. Enderle, C. Mukherjee, B. F{\aa}k \emph{et al}., Europhys. Lett. {\bf 70}, 237 (2005).

\bibitem{Ex2} M. Enderle, B. F\"{a}k, H.-J. Mikeska, R. K. Kremer, A. Prokofiev, and W. Assmus, \prl {\bf 104}, 237207 (2010).

\bibitem{Ex3} M. Mourigal, M. Enderle, B. F{\aa}k, R. K. Kremer, J. M. Law, A. Schneidewind, A. Hiess, and A. Prokofiev, \prl {\bf 109}, 027203 (2012).

\bibitem{Ex4} N. B\"{u}ttgen, P. Kuhns, A. Prokofiev, A. P. Reyes, and L. E. Svistov, \prb {\bf 85}, 214421 (2012).

\bibitem{Ex5} N. B\"{u}ttgen, K. Nawa, T. Fujita, M. Hagiwara, P. Kuhns, A. Prokofiev, A. P. Reyes, L. E. Svistov, K. Yoshimura, and M. Takigawa, \prb {\bf 90}, 134401 (2014).

\bibitem{Ex6} S. Johnston, C. Monney, V. Bisogni \emph{et al}., arxiv:1512.09043.

\bibitem{MF1} S.-W. Cheong and M. Mostovoy, Nat. Mater. {\bf 6}, 13 (2007).

\bibitem{MF2} K. F. Wang, J.-M. Liu, and Z. F. Ren, Adv. Phys. {\bf58}, 321 (2009).

\bibitem{MF3} S. Dong, J.-M. Liu, S.-W. Cheong, and Z. F. Ren, Adv. Phys. {\bf64}, 519 (2015).

\bibitem{Oleg} O. A. Starykh, A. V. Chubukov, and A. G. Abanov, \prb {\bf 74}, 180403 (2006).


\bibitem{Sasha1} A. L. Chernyshev, and M. E. Zhitomirsky, \prl {\bf 97}, 207202 (2006).


\bibitem{Sasha2} A. L. Chernyshev, and M. E. Zhitomirsky, \prb {\bf 79}, 144416 (2009).


\bibitem{Sasha3} M. E. Zhitomirsky, and A. L. Chernyshev, \rmp {\bf 85}, 219 (2013).


\bibitem{IC} S. Nishimoto, S.-L. Drechsler, R. O. Kuzian, J. van den Brink, J. Richter, W. E. A. Lorenz, Y. Skourski, R. Klingeler, and B. B\"{u}chner, \prl {\bf 107}, 097201 (2011).

\bibitem{AG1} S.-L. Drechsler, S. Nishimoto, R. O. Kuzian, J. M¡äalek, W. E. A. Lorenz, J. Richter, J. van den Brink, M. Schmitt, and H. Rosner, \prl {\bf 106}, 219701 (2011).


\bibitem{AG2} M. Enderle, B. F{\aa}k, H.-J. Mikeska and R. K. Kremer, \prl {\bf 106}, 219702 (2011).


\bibitem{AG3} S. Nishimoto, S.-L. Drechsler, R. Kuzian, J. Richter, J. Malek, M. Schmitt, J. van den Brink, and H. Rosner, Europhys. Lett. {\bf 98}, 37007 (2012).


\bibitem{ME} Z. Z. Du, H. M. Liu, Y. L. Xie, Q. H. Wang and J.-M. Liu, \prb {\bf 92}, 214409 (2015).


\bibitem{MWC} N. D. Mermin and H. Wagner, \prl {\bf 17}, 1133 (1966); S. Coleman, Commun. Math. Phys. {\bf 31}, 259 (1973).


\bibitem{QMC} C. Yasuda, S. Todo, K. Hukushima, F. Alet, M. Keller, M. Troyer, and H. Takayama, \prl {\bf 94}, 217201 (2005).


\bibitem{SWT} E. Manousakis, \rmp {\bf 63}, 1 (1991).


\bibitem{HP} T. Holstein, and H. Primakoff, Phys. Rev. {\bf 58}, 1098 (1940).


\bibitem{CCM} R. Zinke, S.-L. Drechsler, and J. Richter, \prb {\bf 79}, 094425 (2009).


\bibitem{OBD1} E. F. Shender, Zh. Eksp. Teor. Fiz. 83, 326 (1982) [Sov. Phys. JETP 56, 178 (1982)].


\bibitem{OBD2} A. L. Chernyshev and M. E. Zhitomirsky, \prl {\bf 113}, 237202 (2014).


\bibitem{CP1} A. K. Kolezhuk, Progress of Theoretical Physics Supplement {\bf 145}, 29 (2002)


\bibitem{CP2} J. Sirker, V. Y. Krivnov, D. V. Dmitriev, A. Herzog, O. Janson, S. Nishimoto, S.-L. Drechsler, and J. Richter, \prb {\bf 84}, 144403 (2011).


\bibitem{Lu} J. Oh, M. D. Le, J. Jeong, J.-H. Lee, H. Woo, W.-Y. Song, T. G. Perring, W. J. L. Buyers, S.-W. Cheong, and J.-G. Park, \prl {\bf 111}, 257202 (2013).

\bibitem{Tr} J. Ma, Y. Kamiya, Tao Hong, H. B. Cao, G. Ehlers, W. Tian, C. D. Batista, Z. L. Dun, H. D. Zhou, and M. Matsuda, \prl {\bf 116}, 087201 (2016).


\bibitem{ESR} H.-A. Krug von Nidda, L. E. Svistov, M. V. Eremin, \emph{et al}., \prb {\bf 65}, 134445 (2002).


\bibitem{Sasha4} M. Mourigal, M. E. Zhitomirsky, and A. L. Chernyshev, \prb {\bf 82}, 144402 (2010).


\bibitem{Sasha5} W. T. Fuhrman, M. Mourigal, M. E. Zhitomirsky, and A. L. Chernyshev, \prb {\bf 85}, 184405 (2012).


\bibitem{Sasha6} M. Mourigal, W. T. Fuhrman, A. L. Chernyshev, and M. E. Zhitomirsky, \prb {\bf 88}, 094407 (2013).


\bibitem{Mahan} G. D. Mahan, \emph{Many-Particle Physics} (Plenum Press, New York, 1990).


\bibitem{NS} G. L. Squires, \emph{Introduction to the Theory of Thermal Neutron Scattering} (Dover, New York, 1996).


\bibitem{LT} C. M. Canali and M. Wallin, \prb {\bf 48}, 3264 (1993).


\bibitem{RIXS} L. J. P. Ament, M. van Veenendaal, T. P. Devereaux, J. P. Hill, and J. van den Brink, \rmp {\bf 83}, 705 (2011).

\end{thebibliography}
\end{document}